# Comparative analysis of financial data differentiation techniques using LSTM neural network

Dominik Stempień[1*] and Janusz Gajda[2]

**Abstract**

We compare traditional approach of computing logarithmic returns with the fractional differencing method and its tempered extension as methods of data preparation before their usage in advanced machine learning models. Differencing parameters are estimated using multiple techniques. The empirical investigation is conducted on data from four major stock indices covering the most recent 10-year period. The set of explanatory variables is additionally extended with technical indicators. The effectiveness of the differencing methods is evaluated using both forecast error metrics and risk-adjusted return trading performance metrics. The findings suggest that fractional differentiation methods provide a suitable data transformation technique, improving the predictive model forecasting performance. Furthermore, the generated predictions appeared to be effective in constructing profitable trading strategies for both individual assets and a portfolio of stock indices. These results underline the importance of appropriate data transformation techniques in financial time series forecasting, supporting the application of memory-preserving techniques.

[1] Faculty of Economic Sciences, University of Warsaw, Długa 44/50, 00-241 Warsaw, Poland
[*] **Corresponding author(s).** E-mail(s): *d.stempien@student.uw.edu.pl*
[2] Department of Statistics and Econometrics, Faculty of Economic Sciences, University of Warsaw, Długa 44/50, 00-241 Warsaw, Poland

## INTRODUCTION

Financial time series forecasting plays a fundamental role in investment decision-making, risk assessment, and the development of trading strategies thus being the huge field of scientific research. Over the years, many novel methodologies and predictive models have been developed in order to enhance the predictive performance. However, due to the low signal-to-noise ratio and substantial uncertainty inherent in financial data, this task remains a significant challenge (De Prado, 2015). One area that has gained attention in recent studies encompasses different data preprocessing techniques before employing the data as input to the predictive model. A promising approach, based on the theory of fractional differentiation by Granger and Joyeux (1980) and Hosking (1981), was introduced by De Prado (2018). This approach ensures that the transformed price series exhibits stationary properties while preserving significantly more of the internal memory (here defined in terms of summability of autocovariance of the time series) than in the case of full order differentiation. In other words, the developed methodology of fractional differentiation retains a substantial amount of information from the original series, which should improve the model's predictive power. As a result, the use of fractionally differentiated price series, instead of logarithmic returns, as input for forecasting models is expected to result in more accurate predictions for unseen observations. Moreover, in this research, some extensions of De Prado's (2018) methodology are derived, which enable the conducting of a more detailed comparative analysis of various data differentiation techniques.

The main objective of this research is to compare the applicability of different data differentiation techniques as inputs to a predictive model. As presented by De Prado (2018), the standard approach of differentiation of order one might remove an excessive amount of memory from the series, which in turn negatively impacts forecasting performance. Because of that, fractional differentiation techniques, which tend to preserve the memory, might constitute a better alternative to traditional approaches. For this reason, four approaches to differentiating the logarithm of the price series are compared: 1) logarithmic returns (standard differentiation of order one), 2) fractionally differentiated series based on De Prado's (2018) method, 3) fractional differentiation based on the ARFIMA model, and 4) fractional differentiation with tempering based on the estimated parameters from the ARTFIMA model. These series, combined with a selected set of technical indicators as features, constitute the inputs to separate Long Short-Term Memory (LSTM) recurrent neural networks.

The predictive quality of the four independent LSTM models is analyzed using performance evaluation metrics. Moreover, the obtained predictions are employed to generate trading signals and corresponding trading strategies. This extension enables assessing the practical applicability of the constructed models in an investment context. The underlying hypothesis of the study states that memory-preserving data differentiation techniques, that is, fractional differencing, constitute a more appropriate method for processing financial time-series data. This should be demonstrated through more accurate predictions and improved values of trading performance indicators for the LSTM models that were based on the fractionally differentiated inputs. Confirmation of the hypothesis would be in line with the findings of Gajda and Walasek (2020). In summary, the following individual hypotheses were formulated in the study:

- *H1:* Memory-preserving data differentiation techniques, represented by fractional differencing, constitute a more appropriate input to the predictive LSTM models compared to logarithmic returns, as indicated by forecasting accuracy metrics.
- *H2:* Memory-preserving data differentiation techniques, represented by fractional differencing, constitute a more appropriate input to the predictive LSTM models compared to logarithmic returns, as indicated by trading performance metrics.
- *H3:* The novel technique of fractional differentiation with tempering might offer an advantage over both logarithmic returns and standard fractional differencing.
- *H4:* The best-performing technique of data differentiation manages to outperform a benchmark *Buy & Hold* approach, as measured by risk-adjusted return metrics.

The empirical investigation of the stated hypothesis is conducted using four datasets comprising major stock indices: S&P 500 (USA), WIG20 (Poland), DAX (Germany), and Nikkei 225 (Japan). The data covers a 10-year period from 01.01.2014 to 31.12.2023, with the first 7 years used as the training set and the final 3 years reserved for the out-of-sample testing set. In addition, the last two years of the training set are separated to form the validation set for the hyperparameter tuning process and model evaluation during training. This data division ensures the robust process of model training, optimal hyperparameter selection, and model evaluation on unseen data.

The obtained findings suggest that fractional differentiation techniques outperform both traditional differentiation of order one and the novel method of fractional differentiation with tempering. The LSTM models utilizing fractionally differentiated price series were characterized by more accurate predictions across the majority of assets. Furthermore, trading strategies based on the signals generated from these models' forecasts achieved superior values of trading performance metrics, particularly in terms of risk-adjusted return indicators. This observation was particularly true in trading applications based on the portfolio of assets rather than individual market indices. However, achieved results indicate that no single technique of fractional differentiation is universally superior. Depending on the asset's specific properties, the most appropriate technique of input transformation alternates between fractionally differentiated series based on the optimal differencing operator derived using De Prado's (2018) method and fractionally differentiated series based on the estimated differencing parameter from the ARFIMA model.

It is important to highlight that, this study contains a thorough comparative analysis of different approaches to time-series data differentiation. Not only is the optimal fractional differencing parameter, based on De Prado's (2018) approach, evaluated, but also the differencing operator from the estimated ARFIMA model. Additionally, the parameters of the ARTFIMA model are estimated and employed to perform tempered fractional differentiation. These phases, which encompass the application of ARFIMA and ARFTIMA model parameters for data differentiation, represent a novel approach that, to the author's knowledge, has not been tested in the relevant literature. This research is based on the most recent data from four widely traded national stock indices. The objective is to evaluate whether the compared techniques constitute a viable option for modelling the current dynamics of different financial markets. Moreover, the LSTM model, which represents a suitable technique for processing sequential data, is applied to provide predictions. The choice of this type of recurrent neural network was driven by the desire to compare different differentiation techniques with a model frequently employed in financial time-series forecasting literature. Finally, the focus of this study goes beyond forecasting only future values of the dependent variable. Equally important is the practical aspect of the generated predictions. For this reason, two types of trading strategies are developed, and their applicability is backtested using historical data. This approach provides a comprehensive evaluation of the usability of the constructed models.

The remaining part of the paper is structured as follows: Section I provides a review of the relevant literature. Special emphasis is placed on the examples of research that employed LSTM neural networks in forecasting and trading applications. A few examples of De Prado's (2018) fractional differentiation method in finance are also discussed. Section II covers the wide variety of employed methodologies, with special attention to the mathematical derivation of the fractional differencing operator. It also introduces the theory of the ARFIMA and ARTFIMA models and their practical application. Additional topics covered in this section include: the computation of technical indicators, a description of performance evaluation metrics, and the logic of the implemented trading strategy. Section III is divided into two segments. Firstly, the datasets applied in the research are described, and the rationale behind the division into the training, validation, and testing data is discussed. In the second part, the estimated values of the differencing parameters are presented. Special emphasis is placed on the clear graphical comparison of the different differentiated price series. Section IV contains a detailed description of the obtained results, with particular attention given to forecasting accuracy and trading performance metrics. Conclusions concerning the practical applicability of the compared differentiation techniques are also discussed. The last segment of the thesis briefly summarizes the research procedure and acquired findings. Moreover, it highlights its contributions to the literature on fractional differencing in the context of financial time-series prediction.

# SECTION I

# Literature review

## 1.1. Approaches to financial data forecasting in light of the Efficient Market Hypothesis

Fama (1970; 1991) presented a systematic theory of efficient capital markets in a series of articles. Efficient Market Hypothesis (EMH) states that prices of financial assets fully reflect all available information. There are three forms of market efficiency described: *weak form*, *semi-strong form,* and *strong form*. All these forms differ from each other in how the set of available information is defined. In the case of *weak-form* market efficiency, the set of information is equal to the historical prices of the asset. *Semi-strong form* further extends the set of information by including publicly available information (for example, fundamental data concerning the company). In a *strong form* of EMH, the set of information additionally includes a group of investors having monopolistic access to non-public data. So, the market might be described as efficient if a given set of information is fully reflected in asset prices. Due to its simplicity, the largest volume of scientific literature addressed the topic of empirical verification of market efficiency in the *weak form.* Acceptance of EMH in *weak form* signifies that technical analysis techniques (analyzing past price movements to forecast their future values) are characterized by no predictive power (Malkiel, 2003). In other words, based only on historical prices, it should not be possible to create a profitable trading strategy that would manage to outperform the market systematically. Fama (1970) performed a literature review focused on the verification of EMH in *weak form*. The obtained conclusions present no evidence to reject the statement that markets are characterized by this category of efficiency. However, up to this day, there is a vast body of literature focused on the empirical testing of market efficiency for different groups of assets. The achieved findings are rather mixed, with some papers supporting the EMH in *weak form* while others violate it (Borges, 2010; Dong et al., 2013; Erdas, 2019; López-Martín et al., 2021; Naseer and Bin Tariq, 2015; Sewell, 2012).

Since the validity of the Efficient Market Hypothesis remains unsolved, there is a broad literature focused on empirical forecasting of financial markets. The traditional approach to stock market analysis is represented by econometric and statistical techniques. This category of models can identify linear dependencies and often requires data to be stationary and normally distributed (Shah et al., 2019). Exponential smoothing models, regression models, and the Autoregressive Integrated Moving Average (ARIMA) model constitute an example of econometric techniques applied to time-series data (Shobana and Umamaheswari, 2021).

Particularly commonly employed is the ARIMA model, which can identify both the momentum and mean reversion effects (Auto-Regressive – AR part) and shock effects (Moving Average – MA part) present in the analyzed series (Shah et al., 2019). ARIMA model is used in the prediction of different types of financial instruments, including market indices (Vo and Ślepaczuk, 2022), individual stocks (Ariyo et al., 2014), currencies (Kumar, 2010), or cryptocurrencies (Wirawan et al., 2019).

Another approach to financial time-series forecasting appeared with the rise in popularity of machine learning models. Machine learning methods may be characterized as data-driven models that can detect both linear and nonlinear dependencies in data, as well as discover complex relationships between dependent variables (Hsu et al., 2016). Since financial data often exhibits described properties, machine learning provided an alternative to the econometric approach. Hsu et al. (2016) performed a comprehensive comparison of different financial market forecasting frameworks. The study was based on 34 national stock market indices from all over the world. One of the aspects of the research was focused on the evaluation of econometric and machine learning approaches. Derived results suggest that machine learning models outperform traditional techniques in terms of prediction accuracy and trading performance. Similar findings have also been achieved in other studies (Akyildirim et al., 2023; Chlebus et al., 2021; Ince and Trafalis, 2008).

In the recent years, due to advances in computational power, significant popularity among forecasting researchers was gained by deep learning models (Bustos and Pomares-Quimbaya, 2020). Deep Neural Networks (DNN) constitute an extension of conventional neural network models. The development of simple neural networks was inspired by the functioning of the human brain, which is constructed out of neurons. Computation operations inside the neuron are based on the weighted sum of its inputs, which are then transformed with nonlinear functions. This process is performed by multiple neurons stacked together into a so-called hidden layer, which takes its inputs from the input layer and generates outputs to the output layer (Sze et al., 2017). DNN expand upon traditional neural networks by incorporating a greater number of hidden layers, which enables automatic feature extraction and transformation (Rouf et al., 2021). A bigger number of hidden layers facilitates more complex and abstract learning, which potentially leads to a better approximation of the underlying function (Sze et al., 2017).. As a result, DNN models find application in many areas of scientific research, managing to outperform other approaches (Schmidhuber, 2015).

Due to the presented advantages of Deep Neural Networks, their application was also popularized in the area of financial markets forecasting. Rouf et al. (2021) analyzed a series of articles from the years 2011-2021 that were focused on stock market prediction with the use of machine learning models. The most popular model in this period was Support Vector Machines (SVM), followed by Artificial Neural Networks (ANN) and DNN. However, the authors underlined that neural network-based models are characterized by more precise predictions. Analogically, Atsalakis and Valavanis (2009) conducted a thorough review of over 100 articles focused on the topic of financial assets forecasting with the use of soft computing methods. The obtained results indicate that advanced neural network architectures usually managed to outperform conventional models both in terms of forecasting accuracy and trading performance. However, the phase of model construction poses a challenge that may significantly influence the results. One category of DNN models, which started gaining increasing recognition, is the Long Short-Term Memory (LSTM) network, which constitutes a recurrent version of DNN (Shah et al., 2019). Due to this tendency and the fact that the LSTM model is employed in this research, the following part of the literature review is focused solely on the examples of applying LSTM networks to financial data forecasting.

## 1.2. Long Short-Term Memory model in the financial forecasting literature

Prediction of financial assets is mainly focused on two related tasks: forecasting future values of the series and evaluating the obtained results with machine learning performance metrics (RMSE, MAE, MAPE, Accuracy, etc.), and employing generated predictions to create a data-based trading strategy. While the first approach is focused rather on the predictive performance of the applied model, the second one aims for a practical assessment of predictions by analyzing the profitability of the strategy on historical data. In the first part of the LSTM-oriented literature review, an emphasis is put on forecasting performance of the LSTM network and its comparison to other models. Subsequently, examples of articles employing LSTM in trading applications are discussed.

Chlebus et al. (2021) performed an extended comparative study of different econometric and machine learning approaches. The investigation was conducted using Nvidia Corporation returns data for the period from 2012 to 2018. Constructed models were provided with different categories of data, including: OHLC (Open, High, Low, Close) price data for Nvidia, other companies, and US market indices; Nvidia's fundamental data, technical indicators based on Nvidia's price data, and behavioral data acquired from Google. Obtained results, based on RMSE, MAE, and MedAE metrics, indicate that the LSTM network was among the best-performing models. Only the SVM model managed to consistently outperform the LSTM.

Another study based on individual stock data was presented by Chen et al. (2015). In this case, the regression approach was replaced by a multiclass classification task. The dataset covered a period of 25 years for over 3000 individual stocks in the Chinese market. The author's findings highlight two conclusions: 1) The LSTM model constitutes the appropriate technique for financial assets forecasting; 2) Using additional features beyond the closing price leads to enhanced model accuracy. Di Persio and Honchar (2017) conducted a comparative study using different types of recurrent neural networks. Basic Recurrent Neural Network (RNN), LSTM, and Gated Recurrent Unit (GRU) models were used to forecast movements of Google stock price data on different time horizons. The LSTM model achieved the highest accuracy, reaching 72% in the case of 5-day ahead prediction.

Fischer and Kraus (2017) undertook the study using individual stock data comprising the S&P 500 index from December 1989 to September 2015. The authors compared several models, including LSTM, DNN, Random Forest, and Logistic Regression. According to Diebold and Mariano's (1995) test, the LSTM model predictions demonstrated superior performance to other non-recurrent alternatives. The authors concluded that the LSTM network constitutes an inherently appropriate method for financial time-series forecasting. Nelson et al. (2017) proposed a comprehensive comparative study of various machine learning models on the example of Brazilian stock price data. The models were fed with exponentially smoothed price data and a set of 175 technical indicators. Data granularity was set to 15 minutes, and the models' task was to predict price directional movement within this interval. In most cases, the LSTM network managed to achieve the most accurate forecasts in comparison to the chosen benchmarks: Random Forest and Multilayer Perceptron network. Moreover, the superior performance of the LSTM model was also confirmed with the Kruskal-Wallis test, which is a non-parametric statistical test used to compare values of multiple samples. Another study employing the LSTM network to forecast the NIFTY 50 stock index was conducted by Roondiwala et al. (2017). The derived conclusions underline the importance of selecting appropriate network hyperparameters and explanatory variables to ensure high performance.

Cryptocurrencies constitute another category of financial assets for which LSTM networks demonstrate promising results. McNally et al. (2018) selected Bitcoin price data intending to compare three models: ARIMA, simple RNN, and LSTM network. The analysis revealed that, in terms of Accuracy and RMSE metrics, both recurrent networks managed to achieve significantly better results than the econometric ARIMA model. Also, the LSTM slightly outperformed its simpler equivalent. However, the authors emphasized the necessity of careful hyperparameter tuning, particularly the dropout rate, to avoid overfitting. Four other cryptocurrencies, including Ethereum, were selected by Ammer and Aldhyani (2022). Features included OHLC price data and volume of the individual cryptocurrency traded during the day. The LSTM model was trained to generate predictions for two time horizons: one day ahead and a long-term forecast over a period of 180 days. The findings provide evidence that the LSTM model serves as an appropriate tool for forecasting cryptocurrency data. On the other hand, Hansun et al. (2022) demonstrated rather contrary findings compared to previous studies. In their approach, the multivariate recurrent models (LSTM, Bidirectional LSTM, and GRU), based on OHLC and Volume data, were applied to forecast the prices of 5 different cryptocurrencies. The findings suggest that the two alternatives to the LSTM network achieved better results. Moreover, the LSTM model results demonstrated greater variation across the assets. However, it should be noted that the hyperparameter tuning process was skipped in the study. Instead, the architecture of the networks was chosen based on the authors' expert knowledge.

LSTM networks are also successfully implemented as a part of hybrid models. The hybrid methodology combines several models with different characteristics in order to better capture underlying patterns present in data (Zhang, 2003). It is based on the assumption that merging diverse individual models allows for obtaining more accurate forecasts. Bukhari et al. (2020) constructed a hybrid ARFIMA-LSTM recurrent network to predict daily price data of Fauji Fertilizer Company. After data preprocessing, the ARFIMA model was fitted to the price series, resulting in obtaining the predictions and residuals. The goal of this step was based on the assumption that the ARFIMA model can extract linear patterns from data. In the next stage, the nonlinear residuals were passed to the LSTM model, which, with the use of additional features, was used to generate another set of forecasts. In the final stage, predictions of ARFIMA and LSTM models were combined, and the final predicted values were achieved. The obtained results indicate that the hybrid ARFIMA-LSTM model managed to outperform its single components: ARFIMA and LSTM models. This conclusion was based on the minimum recorded values of RMSE, MAE, and MAPE metrics. Analogical methodology was also implemented in order to create a hybrid ARIMA-LSTM model for S&P 500 stock index data (Kulshreshtha and Vijayalakshmi, 2020). Empirical findings suggest that the hybrid methodology represents a better alternative to the traditional ARIMA time-series model.

Another framework, consisting of several different methods, was described by Bao et al. (2017). This approach was composed of three independent stages. Firstly, the discrete wavelet transformation was applied to the data in order to denoise it. Next, the denoised data was used as the input for the Stacked Autoencoder (SAE) neural network. SAE constitutes a deep learning model that is trained in an unsupervised manner to extract high-level features and reduce dimensionality. Finally, the LSTM model, trained with denoised and extracted features, was applied to generate one-day-ahead predictions. An empirical investigation of the proposed architecture was performed using multiple worldwide market indices as examples. The results indicated that the employed framework managed to generate more accurate predictions in comparison to an individual LSTM model and a simple RNN.

As mentioned earlier, forecasts of predictive models also find applications in the construction of trading strategies. The next paragraphs briefly describe several studies that employed the LSTM model in this task. Kijewski et al. (2024) conducted a thorough investigation into the profitability of various trading strategies over a period of more than 25 years using S&P 500 price data. Some of the trading strategies were based on the predictive model (ARIMA, LSTM), while the others relied on technical and fundamental factors. The findings suggested a few conclusions. Firstly, the LSTM-based trading strategy was among the best-performing approaches, particularly it managed to significantly outperform the ARIMA model. Secondly, a strategy relying on combining signals from all the techniques doubled the profits of the *Buy & Hold* baseline. Lastly, the sensitivity analysis demonstrated that the LSTM is not robust to changes in parameters. This fact underlines the necessity of an appropriate hyperparameter tuning procedure during the LSTM network training process.

Viéitez et al. (2024) decided to create investment strategies for cryptocurrency Ethereum using multiple sources of data: Ethereum price data, Google statistics, sentiment analysis information, and price data of the assets related to Ethereum. Constructed trading strategies were based on the forecasts of two types of recurrent neural networks: LSTM and GRU. Moreover, independent approaches were constructed for the 1-day, 7-day, and 15-day ahead forecasting. Achieved return and risk metrics suggested that utilizing these models enables the construction of high-performing strategies. In addition, the potential advantage of the LSTM model or the selected time horizon depended on periods of fluctuating volatility in asset prices. LSTM network is also successfully implemented as part of hybrid models in investment applications. Kashif and Ślepaczuk (2025) constructed the hybrid LSTM-ARIMA model to validate its performance across three market indices. The hybrid methodology managed to achieve the highest values of risk-adjusted return metrics for all the assets in comparison to the *Buy & Hold* approach and signals generated by individual ARIMA and LSTM models. However, sensitivity analysis revealed that LSTM-based approaches are not characterized by robustness to changes in hyperparameter values.

It is worth mentioning that the LSTM network also finds applications in novel approaches to algorithmic investment strategies. That is the case of the innovative Mean Absolute Directional Loss (MADL) loss function, which replaces standard loss functions during the LSTM training and validation procedures (Michańków et al., 2022; Michańków et al., 2024). Common loss functions are focused mainly on comparing observed values with predictions, deriving the model's performance based on these values. However, in trading applications, it is crucial to correctly predict directional movement of the asset price, with particularly strong performance when these movements are significant. As a solution, the MADL loss function provides information about both the sign and the magnitude of the potential profit/loss by comparing the actual return with its predicted equivalent. Empirical investigations of MADL applicability to the LSTM model were performed using Bitcoin with Crude Oil price data (Michańków et al., 2024) and Bitcoin with S&P 500 equity index (Michańków et al., 2022). Achieved results, according to risk-adjusted return performance metrics, provided promising conclusions regarding the adaptability of the LSTM model to the new types of loss functions.

### 1.3. Examples of fractional differentiation methodology in financial asset prediction literature

The main inspiration for this research was the study by Gajda and Walasek (2020), which thoroughly investigated the applicability of fractional differentiation in financial time-series forecasting. The analysis was conducted with the use of four worldwide equity indices: S&P 500, WIG20, DAX, and Nikkei 225. The study period covered over 10 years of data from 01.06.2010 to 30.06.2020. For forecasting purposes, the authors decided to employ a feedforward Artificial Neural Network (ANN) with OHLC price data used as the features. Moreover, the data was divided into training and testing sets, while the performance of the predictive model was evaluated with RMSE and MAE metrics. However, the main novelty of the study was the application of fractional differentiation methodology by De Prado (2018). The authors computed the optimal values of the differencing operator for each asset independently and then applied them to derive the fractionally differenced series. The empirical part of the study encompassed a comparison of the predictive performance of constructed networks using standard differenced and fractionally differenced price series, separately. The final findings indicated that fractional differentiation resulted in more accurate predictions for each asset. At the end, it is worth noting that although the methodology of this paper significantly extends the described approach by Gajda and Walasek (2020), the primary idea was taken from that research.

Despite the fact that De Prado's (2018) methodology of fractional differentiation appears to constitute an appropriate technique for financial data transformation, there are few examples of its application in empirical studies. Bieganowski and Ślepaczuk (2024) applied this method, however, in a different manner than described above. The main goal of the research was to generate trading signals using a predictive model for three different cryptocurrencies. The explanatory variables consisted of six time series, including unemployment data and major commodities prices. Instead of applying fractional differencing to the target data (cryptocurrency prices), the authors decided to apply this technique to the features to ensure their stationarity. This operation, combined with other methods performed in the study, resulted in the investment strategy that outperformed the *Buy & Hold* baseline.

# SECTION II

# Methodology

## 2.1. Long Short-Term Memory model

Time-series processes present a unique type of data in which past observations may influence current and future values. That type of dependency in data required developing a more advanced architecture of a deep learning model, which would extend the described earlier simple feedforward neural networks (Kijewski et al., 2024). For this reason, the Recurrent Neural Network (RNN) model was introduced. RNN expands traditional neural networks by including a hidden state, which enables processing sequential information of a time-series (Bao et al., 2017). As a result, RNN demonstrates an ability to learn from past observations, make use of the current state of a process, and utilize this information to provide predictions (Di Persio and Honchar, 2017). Because of those advances, RNN found applications in many areas of research, including language learning, music, electricity load forecasting, and prediction of financial assets (Medsker and Jain, 2001). However, RNN turned out to be incapable of learning long-term dependencies due to insufficient weight changes during the backpropagation algorithm, referred to as the vanishing gradient problem (Hochreiter, 1998). In order to avoid shortcomings of traditional recurrent neural networks, the Long Short-Term Memory (LSTM) model was introduced by Hochreiter and Schmidhuber (1997), which was further developed and evaluated in the following years (Gers et al., 2000; Gers et al., 2002; Greff et al., 2016).

LSTM network belongs to a category of recurrent neural networks however, it solves the vanishing gradient problem. As a result, the LSTM model possesses the capability of learning long-term dependencies that are present in data (McNally et al., 2018). The advantages of the LSTM network derive from its more advanced architecture, which extends traditional RNN. An LSTM network is composed of the input layer, one or more hidden layers, and the output layer. A single hidden layer of an LSTM network is composed of stacked in parallel memory cells, unlike traditional neurons. Memory cells are responsible for storing information and updating it with the use of input, forget, and output gates (Bouteska et al. 2024). Each of the gates completes different tasks: forget gate decides which information is discarded from the cell state, input gate decides which new information is appended to the cell state, and output gate decides which information presents the output of the cell state (Fischer and Kraus, 2018). Memory cell in its internal calculations makes use of sigmoid activation function $sigmoid(x) = \frac{1}{(1+e^{-x})}$ and hyperbolic tangent (tanh) activation function $tanh(x) = \frac{e^x - e^{-x}}{e^x + e^{-x}}$.
Picture 1 presents the structure and flow of the information inside a single memory cell.

**Picture 1.** Architecture of the LSTM network memory cell

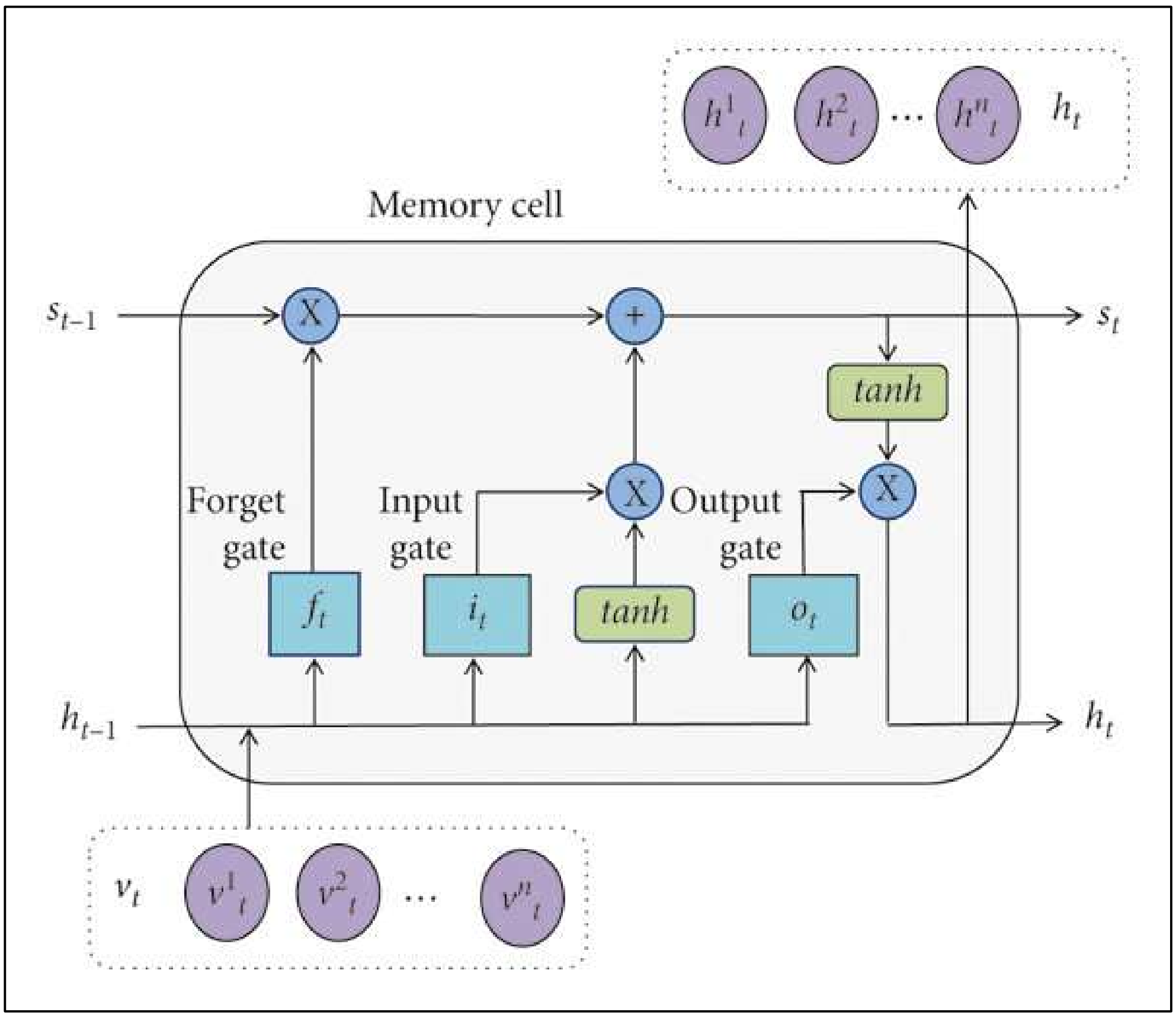

Source: Yang et al. (2020), https://www.researchgate.net/figure/The-structure-of-LSTM-memory-cell_fig5_342998863

The following formulas, based on Yang et al. (2020) and Fischer and Kraus (2018), describe the internal mechanism of the LSTM memory cell:

$$f_t = sigmoid(W_{fv}v_t + W_{fh}h_{t-1} + b_f), \quad [1]$$

$$\tilde{s_t} = tanh(W_{\tilde{s}v}v_t + W_{\tilde{s}h}h_{t-1} + b_{\tilde{s}}), \quad [2]$$

$$i_t = sigmoid(W_{iv}v_t + W_{ih}h_{t-1} + b_i), \quad [3]$$

$$s_t = f_t \odot s_{t-1} + i_t \odot \tilde{s_t}, \quad [4]$$

$$o_t = sigmoid(W_{ov}v_t + W_{oh}h_{t-1} + b_o), \quad [5]$$

$$h_t = o_t \odot tanh(s_t), \quad [6]$$

where:

$v_t$ – input vector at time $t$

$W_{fv}, W_{fh}, W_{\tilde{s}v}, W_{\tilde{s}h}, W_{iv}, W_{ih}, W_{ov}, W_{oh}$ - weight matrices

$b_f, b_{\tilde{s}}, b_i, b_o$ – bias vectors

$f_t$ – forget gate at time $t$

$i_t$ – input gate at time $t$

$o_t$ – output gate at time $t$

$\tilde{s_t}$ – distorted input to memory cell at time $t$

$s_t$ – memory cell at time $t$

$h_t$ – output of the memory cell at time $t$

$\odot$ – elementwise production operator.

At the beginning, LSTM memory cell decides which information from previous cell state $s_{t-1}$ is erased from the memory. This operation is performed on the output of the memory cell at time $t - 1$ $h_{t-1}$ and vector of new input $v_t$ with the use of the forget gate, whose values range from 0 (forget all information) to 1 (remember all information). The next two equations are responsible for determining which new information will update the cell state $s_t$. Consequently, candidate values $\tilde{s_t}$ are computed with *tanh* activation function and activation value of input gate $i_t$ is obtained. Then, new value of cell state $s_t$ is calculated based on how much memory from the previous cell state is retained $f_t \odot s_{t-1}$ and to what extend the cell state is updated $i_t \odot \tilde{s_t}$. In the end, the output of memory cell $h_t$ is finally obtained, representing a filtered version of the calculated cell state $s_t$. This operation is achieved by multiplying activation value of the output gate $o_t$ (values between 0 and 1) and hyperbolic tangent function of cell state $tanh(s_t)$ (values between -1 and 1). In this way, the flow of information inside the LSTM memory cell is performed, where the cell state $s_t$ decides which information is permitted to flow through the network and which is discarded. This property appears to be significantly useful in sequential data processing. During the learning process the weights matrices $W_{fv}, W_{fh}, W_{\tilde{s}v}, W_{\tilde{s}h}, W_{iv}, W_{ih}, W_{ov}, W_{oh}$ and bias vectors $b_f, b_{\tilde{s}}, b_i, b_o$ are updated utilizing Backpropagation Through Time (BPTT) algorithm (Greff et al., 2016). The description of the internal information flow inside a memory cell was inspired by Olah (2015) and Fischer and Kraus (2018).

## 2.2. Fractional differentiation

Many econometric time-series models and supervised machine learning algorithms require data stationarity in order to correctly forecast unobserved values. A stationary time-series is characterized by the fact that it presents no dependency on the specific moment of time in which the series is observed (Hyndman and Athanasopoulos, 2018). In other words, its mean, variance, and covariance should be stable for all the observations in time (Granger and Newbold, 1974). One of the simplest methods of obtaining data stationarity is performed by differencing up to a certain integer order (Newbold, 1975). According to De Prado (2018), assume that variable $X$ represents a non-stationary time-series, with values observed at time $t$:

$$X = \{X_t, X_{t-1}, X_{t-2}, X_{t-3}, \dots, X_{t-k}, \dots\}. \quad [7]$$

In order to transform the series into a stationary one, a differencing transformation must be applied. This procedure is accomplished with the use of time-series operators: $B$, which is a lag operator, and $\Delta$, which represents a difference operator (Harvey, 1990). A lag operator, also called a backshift operator, can be defined as:

$$BX_t = X_{t-1}, \quad [8]$$

$$BX_{t-1} = X_{t-2}, \quad [9]$$

and in general for $k \geq 0$ (Harvey, 1990):

$$B^k X_t = X_{t-k}. \quad [10]$$

In accordance, a difference operator might be defined as:

$$\Delta = 1 - \mathrm{B}, \quad [11]$$

$$\Delta X_t = (1 - \mathrm{B})X_t = X_t - X_{t-1}. \quad [12]$$

The presented transformation is an example of a first-difference operation, which performs a differencing of order one in order to obtain a stationary time-series. In the case when transformed series still do not present stationary properties and fail to pass statistical tests, one might apply a first-difference operator once again (Harvey, 1990):

$$\Delta^2 = (1 - \mathrm{B})^2 = 1 - 2\mathrm{B} + \mathrm{B}^2, \quad [13]$$

$$\Delta^2 X_t = (1 - \mathrm{B})^2 X_t = X_t - 2X_{t-1} + X_{t-2.} \quad [14]$$

In general, applying a first-order differencing operator $d$ times results in a series that is differenced to order $d$ (Harvey, 1990):

$$\Delta^d = (1 - \mathrm{B})^d, \quad [15]$$

$$\Delta^d X_t = (1 - \mathrm{B})^d X_t. \quad [16]$$

Ultimately, the described operations lead to receiving a series with the desired stationary characteristics. However, time-series processes represent a specific kind of data that might preserve memory in a way that current observation reflects the history of past observations. Memory in time-series data might be defined as the presence of statistically significant dependencies between observations at different time lags (Yajima, 1985). As a result of differencing operations, a substantial amount of this memory is removed, which weakens the series' predictive power (Gajda and Walasek, 2020). De Prado (2018) thoroughly describes this problem by naming it the stationarity vs. memory dilemma. For example, price series of the financial assets are mostly non-stationary but preserve memory. That is, future observations are in some way related to the past ones. On the other hand, returns or logarithmic returns present stationary properties, but the memory is significantly removed. Overdifferencing and cutting an excessive amount of memory can be particularly costly in financial applications, as financial time-series are typically characterized by a high level of noise, making it challenging to extract meaningful signals (De Prado, 2015).

As a result of the stationarity vs. memory dilemma, De Prado (2018) raises a question whether there exists a different method of imposing data stationarity while keeping as much memory as possible. In other words, the challenge is to find an optimal way of financial data transformation that would operate between two extremes of no differentiation and differentiation of order one. For this purpose, De Prado employed the concept of fractional differentiation, which was first introduced by Hosking (1981). He extended it with the use of the binomial theorem in order to enable differencing of non-integer orders. For any nonnegative integer $n$, the following binomial formula holds (De Prado, 2018):

$$(x+y)^n = \sum_{k=0}^{n} \binom{n}{k} x^k y^{n-k} = \sum_{k=0}^{n} \binom{n}{k} x^{n-k} y^k. \quad [17]$$

By substituting $y$ with unity and $d$ being a real number (De Prado, 2018):

$$(1+x)^d = \sum_{k=0}^{\infty} \binom{d}{k} x^k. \quad [18]$$

Due to the use of $d$ as a real number, differentiation of order $d$ presented in Equations [15] and [16] can be expanded by the fractional model in the following way (De Prado, 2018):

$$\begin{aligned}(1-\mathrm{B})^d &= \sum_{k=0}^{\infty} \binom{d}{k} (-B)^k = \sum_{k=0}^{\infty} \frac{\prod_{i=0}^{k-1}(d-i)}{k!} (-B)^k \\ &= \sum_{k=0}^{\infty} (-B)^k \prod_{i=0}^{k-1} \frac{(d-i)}{(k-i)} = 1 - dB + \frac{d(d-1)}{2!} B^2 \\ &- \frac{d(d-1)(d-2)}{3!} B^3 + \cdots + (-B)^k \prod_{i=0}^{k-1} \frac{(d-i)}{k!} + \cdots\end{aligned} \quad [19]$$

$$\begin{aligned}\Delta^d X_t &= (1-B)^d X_t \\ &= X_t - dX_{t-1} + \frac{d(d-1)}{2!} X_{t-2} - \frac{d(d-1)(d-2)}{3!} X_{t-3} + \cdots \\ &+ (-1)^k \prod_{i=0}^{k-1} \frac{(d-i)}{k!} X_{t-k} + \cdots\end{aligned} \quad [20]$$

After performing a fractional differentiation procedure, the weights corresponding to the particular observations are calculated and might be expressed as ω (De Prado, 2018):

$$\omega = \left\{1, -d, \frac{d(d-1)}{2!}, -\frac{d(d-1)(d-2)}{3!}, \ldots, (-1)^k \prod_{i=0}^{k-1} \frac{(d-i)}{k!}, \ldots\right\}. \quad [21]$$

In the case when $d$ is a positive integer, there always exists an observation beyond which the corresponding weights become zero. This occurs because, for $d = i$, the formula $(-1)^k \prod_{i=0}^{k-1} \frac{(d-i)}{k!}$ is equal to zero. It means that after differencing operation memory beyond this point got removed from the series. For example, differentiation of a first order results in the following set of weights $\omega = \{1, -1, 0, 0, \dots\}$. Deducing from the Equation [21], set of weights ω can be formulated in an iterative manner in the following way (De Prado, 2018):

$$\omega_0 = 1, \quad [22]$$

$$\omega_k = -\omega_{k-1} \frac{d-k+1}{k}. \quad [23]$$

Figure 1 presents a comparison of the weights $\omega_k$ corresponding to particular observations $X_{t-k}$ in the series after performing fractional differentiation for different values of $d \in [0, 1]$. In the case of no differentiation, $d = 0$, the only non-zero value corresponds to the current value of a time-series. For $d = 1$, that is when first-order differencing is applied, $\omega_0 = 1$ and $\omega_1 = -1$. The resulting set of weights corresponds to the transformation of the logarithm of the price series into logarithmic returns. In other scenarios, for $d \in (0, 1)$ all weights beyond $\omega_0$ are negative and converge asymptotically to zero at different rates. Non-zero values of these weights indicate that the memory of past observations was preserved (De Prado, 2018). For a complete picture of the impact of parameter $d$ on weights, Figure 2 demonstrates a comparison of the weights $\omega_k$ for $d \geq 1$.

As indicated above, for $d \in (0, 1)$, the weights for all past observations take non-zero values. For practical reasons, this fact poses a challenge in employing fractional differentiation in real-life applications. For this reason, De Prado (2018) proposed conducting fractional differentiation with the use of the fixed-width window method. The procedure involves dropping the past observations for which the modulus of their weights is lower than a certain threshold τ. A new set of weights $\tilde{\omega}$, can be presented mathematically in the following way (De Prado, 2018):

$$\widetilde{\omega_k} = \begin{cases} \omega_k \ if \ |\omega_k| \geq \tau \\ 0 \ if \ |\omega_k| < \tau \end{cases}. \quad [24]$$

**Figure 1.** Comparison of the weights for subsequent lagged observations for $d \in [0, 1]$

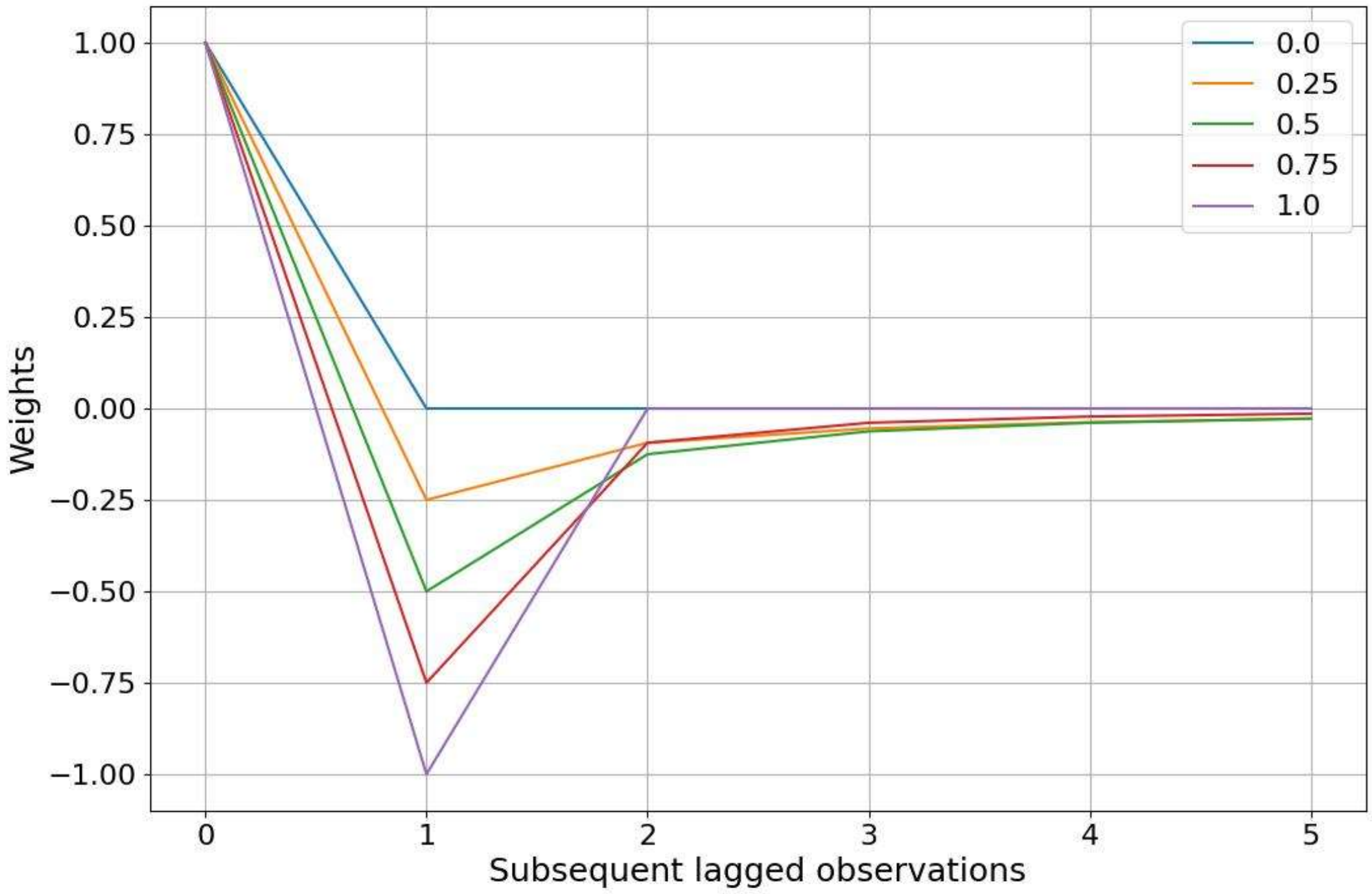

Source: Own elaborations based on De Prado (2018)

The main goal of the described transformations is to obtain a stationary series with maximum memory preservation. In particular, the objective is to find the smallest value of $d^*$ for which the given series presents stationary properties. For this purpose, De Prado (2018) employs an augmented Dickey-Fuller test (ADF test) assuming a 95% confidence level. Rejection of the null hypothesis denotes that there is no presence of a unit root in the data, meaning that the analyzed time-series is already stationary. Applying a fractional differentiation operator with a value greater than a minimum coefficient $d^*$ would result in overdifferencing and removing an excessive amount of memory. Especially, that is the case while applying differentiation of the first order. On an example of price data of almost one hundred most frequently traded future contracts in the world, De Prado (2018) showed that all of them achieved stationarity with $d < 0.6$, while for the vast majority of them, $d < 0.3$ was sufficient. Analogous conclusions were demonstrated by Gajda and Walasek (2020) for main stock indices worldwide. Computed optimal values of the fractional differencing operator oscillated between the values of 0.12 and 0.43.

**Figure 2.** Comparison of the weights for subsequent lagged observations for $d \geq 1$

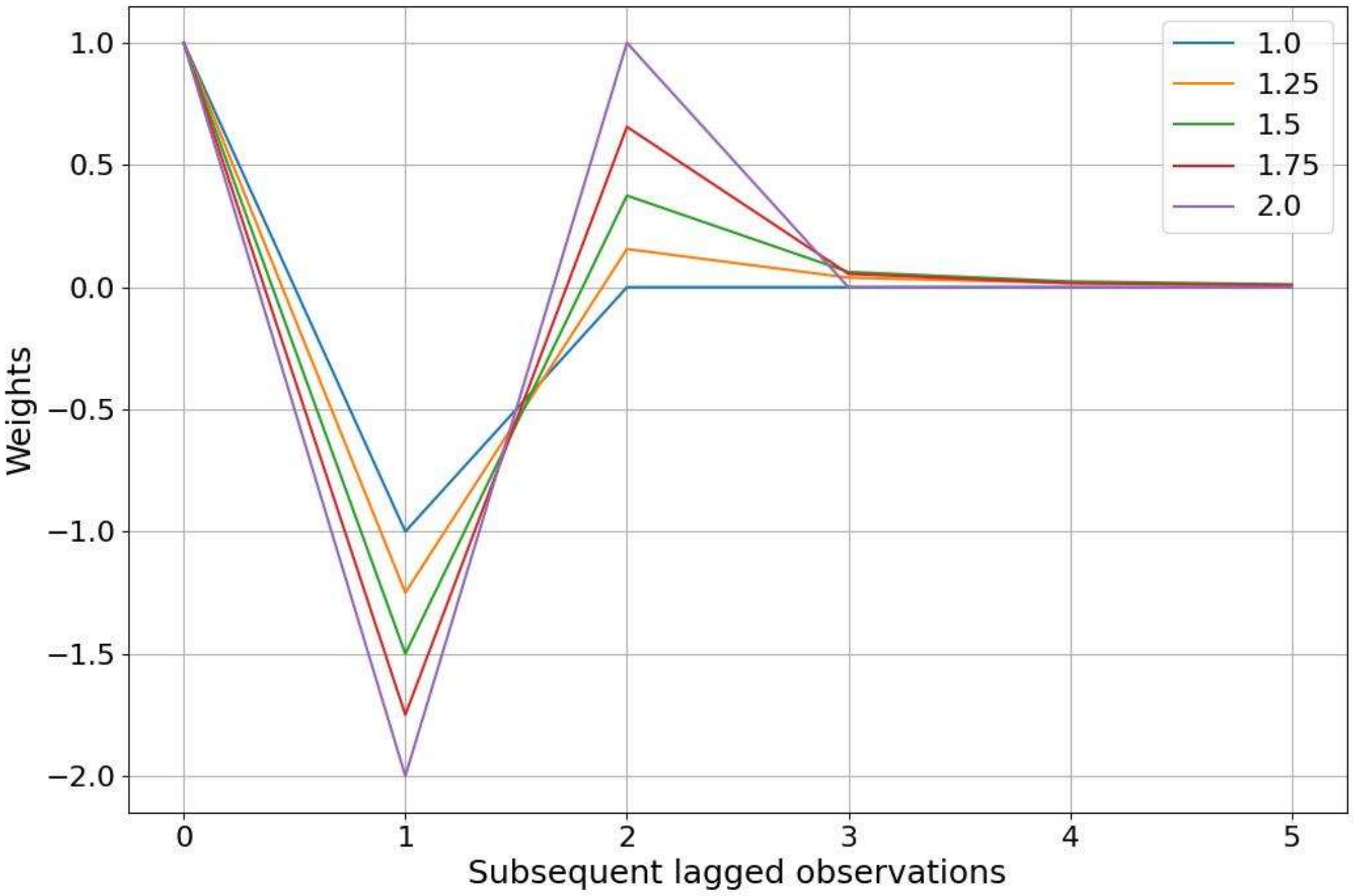

Source: Own elaborations based on De Prado (2018)

### 2.3. Autoregressive Fractionally Integrated Moving Average model

The methodology described by De Prado (2018) was inspired by the Autoregressive Fractionally Integrated Moving Average time-series model (ARFIMA), originally proposed by Granger and Joyeux (1980) and Hosking (1981). One of the main reasons for developing a novel time-series process was an observation that a sample autocorrelation function (ACF) in many empirical datasets is characterized by a slower decline than in the ARIMA model with an integer order of differencing (De Gooijer and Hyndman, 2006). Idea of the ARFIMA$(p, d, q)$ model is mainly based on fractional differencing operator presented in Equation [19]. It constitutes an extension of the ARIMA$(p, d, q)$ model by allowing $d$ parameter to take non-integer values. Based on Hosking (1981), it can be stated that process $X$ follows ARFIMA$(p, d, q)$ model for $d \notin \mathbb{Z}$ if

$$Y_t = \Delta^d X_t = (1 - \mathrm{B})^d X_t \qquad [25]$$

follows the ARMA$(p, q)$ model:

$$Y_t - \sum_{j=1}^{p} \phi_j Y_{t-j} = Z_t - \sum_{i=1}^{q} \theta_i Z_{t-i} \quad [26]$$

where:

$Z$ – sequence of white noise error terms

$\phi_1, \ldots, \phi_p$ – parameters of the autoregressive model of order $p$

$\theta_1, \ldots, \theta_q$ – parameters of the moving average model of order $q$.

Special case of ARFIMA$(p, d, q)$ model is attained if $p = 0$ and $q = 0$. In this situation, the autoregressive and moving average parts of the models are omitted and the process might be represented as ARFIMA$(0, d, 0)$. If $d \in (-0.5, 0.5)$, the process is stationary and invertible. In the most interesting case, in which $d \in (0, 0.5)$, ARFIMA$(0, d, 0)$ is additionally characterized by the presence of long-term memory (Hosking, 1981). According to Granger and Ding (1996), long memory of the series might be described in the following manner: observations of a series exhibit a long-term memory if a shock $\varepsilon_t$ at time $t$ continues to have an impact on future $x_{t+k}$ for a greater horizon $k$ than in the case of ARIMA process. In other words, the ARFIMA model exhibits a hyperbolic decay pattern of autocorrelation between observations while the ARIMA process displays an exponential decay structure which drops significantly faster (Koustas and Serletis, 2005; Oomen, 2001). Granger and Joyeux (1980) underline that this property of the ARFIMA model might be especially useful in the instance of long-term forecasting.

There is a broad body of literature concerning the topic of the presence of long-term memory in financial datasets. Assaf (2006) performed a research focused on the exploration of long memory in four stock markets belonging to the Middle East and North Africa (MENA) region. For this purpose, two statistical tests, focused on detecting long-range dependencies, were employed. The obtained results suggested the presence of long memory in stock index returns for two of four countries. However, long memory in volatility series was confirmed for all the markets. Floros et al. (2007) applied the conditional maximum likelihood method in order to estimate three models: ARFIMA, ARFIMA-GARCH, and ARFIMA-FIGARCH. The study was based on daily returns of the Portuguese PSI20 index. The estimated $d$ parameter was positive and statistically significant at least 5% level across all models. Identified long-term memory suggested that in the period 1993-2006 Portuguese stock market was predictable in the long-horizon. On the other hand, there are also some publications challenging the existence of long-term memory in financial assets. Cheung and Lai (1995) made use of stock market data belonging to 18 countries worldwide. Instead of applying a daily frequency to compute returns, a monthly return series was chosen in order to test for long memory. Moreover, not only were nominal returns calculated, but also real market returns, which were adjusted for the US inflation rate. Depending on the applied statistical test, significant evidence for long-term dependencies was found only for 5 countries. Similar results of no long memory were demonstrated for daily and monthly US stock market returns (Lo, 1991).

At the same time, there are also publications focusing on the empirical forecasting of financial data with the use of long-term memory models. Bhardwaj and Swanson (2005) performed a broad comparison of the efficiency of the ARFIMA model versus other time-series models: random walk (with and without drift), AR, MA, ARMA, ARIMA, and GARCH. Data consisted of several decades of price series for the most important stock market indices, including S&P 500, DAX, and Nikkei 225. Moreover, forecasts were generated for 1-day, 5-day, and 20-day ahead, and different transformations of market returns were compared. Obtained results, based on Diebold and Mariano's (1995) DM test, strongly suggested that the ARFIMA model presents the most effective forecasting properties. This observation even intensified for longer prediction horizons. Similar results presenting good forecasting properties for the ARFIMA model were achieved for the Athens Stock Exchange in Greece (Barkoulas et al., 2000) and stock markets of countries belonging to the MENA region (Assaf, 2006). At the end, it is worth mentioning that fractional models are successfully applied in other fields of research, including biology, telecommunication, and astrophysics (Burnecki and Weron, 2014).

### 2.4. Autoregressive Tempered Fractionally Integrated Moving Average model

A further extension of fractional differencing was proposed by Meerschaert et al. (2015), who introduced a new form of fractional calculus operation that incorporates an exponential factor to temper power laws. Multiplying fractional derivatives by an exponential component results in obtaining tempered fractional derivatives. Compared to standard fractional calculus, tempered fractional calculus offers a more flexible framework for modelling time-series data. In particular, the inclusion of the exponential factor allows for accurate modelling of semi-long dependencies that transition from long-range to short-range correlations (Meerschaert et al., 2015). In other words, this refers to time-series data in which the autocovariance function initially decays slowly but then decays more rapidly at larger lags (Sabzikar et al, 2020). The extended theory of tempered fractional calculus and mathematical derivation of formulas is out of the scope of this work but can be found in Meerschaert et al. (2015) or Sabzikar and Kokoszka (2023).

Tempered fractional calculus extends the fractional calculus approach by incorporating one additional parameter λ. While parameter $d$ is still related to the magnitude of long-range dependencies, the λ parameter controls the rate of decay of these correlations (Sabzikar and Kokoszka, 2023). Analogous to the formulation of integer and fractional difference operators, tempered fractional difference operator is defined as follows, where $d > 0$, $d \notin \mathbb{Z}$ and $\lambda > 0$ (Sabzikar et al., 2019):

$$\Delta^{d,\lambda} = (1 - e^{-\lambda}\mathrm{B})^{d}, \quad [27]$$

$$\Delta^{d,\lambda} X_t = (1 - e^{-\lambda}\mathrm{B})^{d} X_t. \quad [28]$$

Once again, using the binomial theorem and expanding the formulas above:

$$\begin{aligned}(1 - e^{-\lambda}\mathrm{B})^{d} &= \sum_{k=0}^{\infty} \binom{d}{k} (-e^{-\lambda}B)^{k} = \sum_{k=0}^{\infty} \frac{\prod_{i=0}^{k-1}(d-i)}{k!} (-e^{-\lambda}B)^{k} \\ &= \sum_{k=0}^{\infty} (-e^{-\lambda}B)^{k} \prod_{i=0}^{k-1} \frac{(d-i)}{(k-i)} = 1 - e^{-\lambda} dB + e^{-2\lambda} \frac{d(d-1)}{2!} B^{2} \\ &- e^{-3\lambda} \frac{d(d-1)(d-2)}{3!} B^{3} + \cdots + e^{-k\lambda} (-B)^{k} \prod_{i=0}^{k-1} \frac{(d-i)}{k!} + \cdots \end{aligned} \quad [29]$$

$$\begin{aligned}\Delta^{d,\lambda} X_t &= (1 - e^{-\lambda}\mathrm{B})^{d} X_t \\ &= X_t - e^{-\lambda} d X_{t-1} + e^{-2\lambda} \frac{d(d-1)}{2!} X_{t-2} \\ &- e^{-3\lambda} \frac{d(d-1)(d-2)}{3!} X_{t-3} + \cdots + e^{-k\lambda} (-1)^{k} \prod_{i=0}^{k-1} \frac{(d-i)}{k!} X_{t-k} \\ &+ \cdots \end{aligned} \quad [30]$$

Finally, vector of weights $\omega = \{\omega_0, \omega_1, \omega_2, \omega_3, \ldots, \omega_k, \ldots\}$ corresponding to the observations $X = \{X_t, X_{t-1}, X_{t-2}, X_{t-3}, \ldots, X_{t-k}, \ldots\}$ is expressed by (Sabzikar et al., 2019):

$$\omega = \left\{1, -e^{-\lambda}d,\ e^{-2\lambda}\frac{d(d-1)}{2!}, -e^{-3\lambda}\frac{d(d-1)(d-2)}{3!}, \dots, e^{-k\lambda}(-1)^k \prod_{i=0}^{k-1}\frac{(d-i)}{k!}, \dots\right\}, \quad [31]$$

$$\omega_k = e^{-k\lambda}(-1)^k \binom{d}{k}. \quad [32]$$

So, it is clear that for $\lambda = 0$ and $d$ being any positive real non-integer number, computed weights would be equal to the weights obtained by the method of fractional differentiation. Based on the derived formula for tempered fractional difference operator in Equation [29], Meerschaert et al. (2015) introduced an Autoregressive Tempered Fractionally Integrated Moving Average model (ARTFIMA). According to definition presented in Sabzikar et al. (2019), process $X$ follows ARTFIMA$(p, d, \lambda, q)$ model for $d \notin \mathbb{Z}$ and $\lambda > 0$ if

$$Y_t = \Delta^{d,\lambda} X_t = (1 - e^{-\lambda}\mathrm{B})^d X_t \quad [33]$$

follows the ARMA$(p, q)$ model:

$$Y_t - \sum_{j=1}^{p} \phi_j Y_{t-j} = Z_t - \sum_{i=1}^{q} \theta_i Z_{t-i} \quad [34]$$

where:

$Z$ – sequence of white noise error terms

$\phi_1, \dots, \phi_p$ – parameters of the autoregressive model of order $p$

$\theta_1, \dots, \theta_q$ – parameters of the moving average model of order $q$.

ARTFIMA$(p, d, \lambda, q)$ model presents an extension of earlier described ARFIMA$(p, d, q)$ model. Especially, in the case of $\lambda = 0$, both models are identical. Sabzikar et al. (2019) points out three main advantages of applying ARTFIMA over ARFIMA: 1) Unlike the ARFIMA model, ARTFIMA possesses a summable covariance function and its properties can be obtained through standard methods; 2) ARTFIMA model demonstrates a better fit to spectral density data for low frequencies; 3) While ARFIMA model presents stationary properties only for $d \in (-0.5, 0.5)$, owing to $\lambda$ parameter ARTFIMA model is stationary for any $d \notin \mathbb{Z}$. Empirical investigations of the usefulness of ARTFIMA model and its better fit to data comparing to ARFIMA model was additionally proved in the series of studies including:

water velocity data (Meerschaert et al., 2015; Sabzikar et al., 2019), solar-flare data (Kabala and Sabzikar, 2021), adjusted closing stock prices (Sabzikar et al., 2019), and mRNA dataset (Sabzikar et al., 2022). Mentioned simulations were performed with the use of the $R$ programming language package *artfima,* which fits and simulates ARTFIMA models (Sabzikar et al., 2019).

A particular case of ARTFIMA$(p, d, \lambda, q)$ is derived for $p = 0$ and $q = 0$. Obtained ARTFIMA$(0, d, \lambda, 0)$ process comes down to tempered fractional difference operator presented in the Equation [30]. Analogically, ARTFIMA$(0, d, \lambda, 0)$ is characterized by semi-long range dependence structure with long range dependencies for the first lags and covariance function which decays in an exponential way (Sabzikar et al., 2019). The rate of this decline is determined by the λ parameter.

### 2.5. Applying De Prado's (2018) approach for $d$ and λ parameters estimated by ARFIMA and ARTFIMA models

Presented earlier De Prado's (2018) method of fractional differentiation enables obtaining a stationary time-series process for a parameter $d$ that preserves the maximum possible amount of memory. As a result, set of weights $\omega = \{\omega_0, \omega_1, \omega_2, \omega_3, \ldots, \omega_k, \ldots\}$ corresponding to the observations $X = \{X_t, X_{t-1}, X_{t-2}, X_{t-3}, \ldots, X_{t-k}, \ldots\}$ is obtained. From now, parameter $d$ which is calculated in the described way will be represented by parameter $d_{De\ Prado}$ and set of weights as $\widetilde{\omega}_{De\ Prado}$. However, as presented in the section concerning ARFIMA model, parameter $d_{ARFIMA}$ might be also computed by estimating ARFIMA$(0, d, 0)$ model. After applying De Prado (2018) methodology of calculating weights for $d_{ARFIMA}$ parameter (Equation [21]), new set of weights $\widetilde{\omega}_{ARFIMA}$ might be obtained and used further in order to transform observations.

On the other hand, implementing De Prado's (2018) procedure for parameters estimated by the ARTFIMA model requires a more sophisticated approach. After the estimation of ARTFIMA$(0, d, \lambda, 0)$ model, the values of two parameters are computed: $d_{ARTFIMA}$ and $\lambda_{ARTFIMA}$. In order to employ them in the transformation of the price series, the equations derived by De Prado (2018) must be rewritten in a way that acknowledges the existence of an additional parameter. Equation [31] presents the values of weights obtained after applying the fractional tempered difference operator. After analyzing the sequence of consecutive values, it is clear that Equations [22] and [23], which present the formulas for computing the value of weights iteratively, might be transformed in the following way:

$$\omega_0 = 1, \tag{35}$$

$$\omega_k = -\omega_{k-1} \frac{d - k + 1}{k} e^{-\lambda}. \tag{36}$$

This sequential procedure and fixed-width window method from Equation [24] result in the computed set of weights $\tilde{\omega}_{ARTFIMA}$. Consequently, the parameters $d_{De\ Prado}$, $d_{ARFIMA}$ and the set of $d_{ARTFIMA}$ and $\lambda_{ARTFIMA}$ allow for the construction of three distinct sets of weights: $\tilde{\omega}_{De\ Prado}$, $\tilde{\omega}_{ARFIMA}$ and $\tilde{\omega}_{ARTFIMA}$. These weights enable the generation of four differently differentiated time-series: one fully differentiated series (logarithmic returns) – $\Delta X_t$, two fractionally differentiated series – $\Delta^{d_{De\ Prado}} X_t$ and $\Delta^{d_{ARFIMA}} X_t$, and one tempered fractionally differentiated series – $\Delta^{d_{ARTFIMA}, \lambda_{ARTFIMA}} X_t$. To the author's knowledge, the application of the methodology by De Prado (2018) for the parameter $d$ estimated by ARFIMA$(0, d, 0)$ model, and further extended through the incorporation of the fractionally tempered set of weights, in order to generate different inputs for a forecasting model, has not been addressed in the existing literature. This approach presents a novelty in the field of financial data transformation and prediction. Moreover, the main goal of this research is to quantitatively compare described methods of data differentiation by generating distinct sets of predictions with the use of LSTM neural network.

### 2.6. Technical indicators

In order to predict unseen values of forecasted price series, not only are its past values used, but also a set of selected technical indicators. In this research, four different technical analysis techniques are employed: Simple Moving Average (SMA), Relative Strength Index (RSI), Bollinger Bands (BBands), and Moving Average Convergence / Divergence (MACD). Technical indicators represent a set of mathematical formulas used by some traders with the hope of correctly predicting an asset's future price movements (Cocco, 2021). In this research, technical indicators are utilized as the features for constructing LSTM models with the intention that the deep learning model will be able to extract some significant relationships between their values. The section below shortly describes each of the chosen indicators and contains the appropriate mathematical formula:

- Simple Moving Average (SMA) – probably one of the simplest technical indicators used by traders. $SMA_k$ is calculated as a simple unweighted average of prices of last $k$ assets' observations. $SMA_k$ only provides information concerning price series trend.

$$SMA_k = \frac{1}{k} \sum_{i=N-k+1}^{N} p_i \qquad [37]$$

where:

$N$ – number of observations

$k$ – number of past days for which the metric is calculated

$p_i$ – price of the asset at day $i$.

- Bollinger Bands (BBands) – trading bands which further extend the idea of Simple Moving Average. They are graphically represented as two lines: first one is above $SMA_k$ by standard deviation of prices multiplied by factor $d$ ($Upper\ BBand_{k,d}$), and second one is below by the same value ($Lower\ BBand_{k,d}$). Traditionally, $d$ parameter is equal to 2. If price of the financial asset extends one of the bands, it is expected to revert back to the levels determined by Simple Moving Average. Thus, Bollinger Bands generate trading signals for buy or sell depending on which band is crossed (Lento et al., 2007).

$$Upper\ BBand_{k,d} = SMA_k + \sigma_k * d \qquad [38]$$

$$Lower\ BBand_{k,d} = SMA_k - \sigma_k * d \qquad [39]$$

where:

$\sigma_K$ – standard deviation of prices of the asset over a period $k$

$d$ – constant multiplier.

- Relative Strength Index (RSI) – an oscillator with values ranging between 0 and 100 (Kim, 2003). $RSI_k$ measures dynamics of changing asset's prices over a period of last $k$ days (typically $k = 14$). The values above 70 indicate an overbought scenario of the asset and generate a sell signal for traders. On the other hand, values below 30 are understood as an oversold scenario and create a buy signal. Area between these values is considered neutral (Vo and Yost-Bremm, 2020).

$$RSI_k = 100 - \frac{100}{1 + RS_k} \quad [40]$$

where:

$RS_k$ – ratio of average daily gain value to average daily loss value during last $k$ days.

- Moving Average Convergence / Divergence (MACD) – popular technical indicator providing signals concerning trend and momentum of the price series. It contains two sets of information: $MACD_{p,q}$ and $Signal\ Line$. The $MACD_{p,q}$ line is constructed as the difference between short-term ($p$-period) and long-term ($q$-period) exponential moving averages of the asset's prices. On the other hand, $Signal\ Line$ is calculated as exponential moving average of past $r$ $MACD_{p,q}$ values. For traders, situation in which $MACD_{p,q}$ is placed above the $Signal\ Line$ indicates a buy signal. Sell signal arises in the opposite case. Standard values for $p$, $q$ and $r$ parameters are 12, 26, and 9 (Vo and Yost-Bremm, 2020).

$$MACD_{p,q} = EMA_p - EMA_q \quad [41]$$

$$Signal\ Line = EMA_r(MACD) \quad [42]$$

where:

$EMA_k$ – exponential moving average of prices of the asset over a period $k$.

## 2.7. Performance metrics

Forecasting capabilities of constructed LSTM networks based on different price series transformations are evaluated on training, validation, and testing sets with the use of three different performance metrics. This division enables a better understanding of the reliability of the obtained predictions. Chosen metrics present in the following order:

- Root Mean Squared Error (RMSE)

$$RMSE = \sqrt{\frac{\sum_{i=1}^{N}(\widehat{y_i} - y_i)^2}{N}} \quad [43]$$

- Mean Absolute Error (MAE)

$$MAE = \frac{\sum_{i=1}^{N} |\hat{y}_i - y_i|}{N} \qquad [44]$$

- Mean Absolute Percentage Error (MAPE)

$$MAPE = \frac{\sum_{i=1}^{N} \left| \frac{\hat{y}_i - y_i}{y_i} \right|}{N} \qquad [45]$$

where:

$y_i$ – actual value for observation $i$

$\hat{y}_i$ – predicted value for observation $i$

$N$ – number of observations.

## 2.8. Trading strategies

Forecasting financial time-series is characterized by some unique implications in comparison to other types of data. On the one hand, there is some common factor – the goal is to generate predictions that are close to the factual realizations of the series. This aspect is measured by the earlier described performance metrics: RMSE, MAE, and MAPE. However, in a case of financial assets forecasting, the practical aspect of prediction accuracy is equally important. For this reason, obtained predictions should be employed as a foundation for a trading strategy evaluated on the out-of-sample historical market data. In the financial literature, this procedure is referred to as *backtesting*. Backtesting is a historical simulation of market data during which potential gains and losses from the employed trading strategy are calculated over a defined period (Bailey et al., 2014).

In order to avoid a risk of backtest overfitting (Bailey et al., 2016), two relatively simple trading strategies are compared in this paper: $Long - Short$ and $Long - Only$ (Kashif and Ślepaczuk, 2025). $Long - Short$ trading strategy enables to open either a long position (1) or short position (-1). On the other hand, in $Long\ Only$ strategy, only a long position (1) or holding no position (0) is allowed. To thoroughly simulate the functioning of financial markets, the process of opening and closing positions is associated with transaction costs. Moreover, opening a new position or modifying an existing one takes place only if potential gains associated with a trade exceed transaction costs. Process of generating signals for both $Long - Short$ and $Long\ Only$ trading strategies might be formulated in the following way:

$$Long - Short:\begin{cases} Signal = 1\ if\ \widehat{y_{i+1}} > \ y_i * (1 + c) \\ Signal = -1\ if\ \widehat{y_{i+1}} < \ y_i * (1 - c) \end{cases} \quad [46]$$

$$Long\ Only:\begin{cases} Signal = 1\ if\ \widehat{y_{i+1}} > \ y_i * (1 + c) \\ Signal = 0\ if\ \widehat{y_{i+1}} < \ y_i * (1 - c) \end{cases} \quad [47]$$

where:

$y_i$ – actual value for observation $i$

$\widehat{y_{i+1}}$ – predicted value for observation $i + 1$ made at the end of day $i$

$c$ – transaction costs (%).

### 2.9. Trading strategy's performance indicators

In order to thoroughly compare the performance of trading strategies for different data transformation techniques and assets, various trading efficiency metrics are employed. The selected performance indicators underline different aspects of the trading strategy, including its returns, risk, and risk-adjusted returns. The application of trading performance indicators provides a deeper understanding of a trading strategy than the equity line alone. Section below, which was influenced by Ryś and Ślepaczuk (2019) and Gómez and Ślepaczuk (2021), presents selected indicators with their short description and appropriate mathematical formula:

- Annualized rate of return (ARC) – a simple metric informing about the average return of the investment in a yearly interval. ARC presents the mean percentage change in an asset's value in the year-to-year horizon, accounting for compounding of returns. ARC provides only a narrow perspective of investment performance due to the fact that there is no information about the volatility of returns and risk.

$$ARC = \left(\frac{V_{t_2}}{V_{t_1}}\right)^{\frac{1}{D(t_1,t_2)}} - 1 \quad [48]$$

where:

$V_{t_i}$ – value of the asset at time $t_i$

$D(t_1, t_2)$ – number of years between $t_1$ and $t_2$.

- Annualized standard deviation (ASD) – empirical standard deviation of returns expressed on an annual basis (number 252 is used to approximate an average number of trading days during the year). The metric informs about the volatility of the applied trading strategy.

$$ASD = \sqrt{252} * \sqrt{\frac{1}{N-1}\sum_{i=1}^{N}(r_i - \bar{r})^2} \quad [49]$$

where:

$N$ – number of trading days

$r_i$ – assets' percentage return on day $i$

$\bar{r}$ – assets' average daily percentage return.

- Maximum Drawdown (MD) – metric representing the maximum percentage decline in value of the equity line. In other words, it is the biggest drop in the portfolio's value for the applied trading strategy. Important metric from the perspective of an active trader who may be characterized by an aversion to significant declines in portfolio value during strategy lifetime. Small values of the Maximum Drawdown metric signify that the strategy is more resistant to chaotic market dynamics (Pareschi, 2003).

$$MD = sup_{x,y \in \{[t_1, t_2]^2 : x \le y\}} \frac{P_x - P_y}{P_x} \quad [50]$$

where:

$P_t$ – equity line level at time $t$.

- Information Ratio (IR) – metric taking into consideration two aspects of trading strategy: returns and associated risk. It is calculated as the ratio of the annualized rate of return to the annualized standard deviation. Risk-adjusted returns provide a more holistic approach to understand and compare strategy performance versus some benchmark. The high value of the metric indicates that the strategy delivers significant returns in comparison to relatively low volatility.

$$IR = \frac{ARC}{ASD} \qquad [51]$$

- Sortino Ratio (SR) – variation of the described above Information Ratio by replacing annualized standard deviation with annualized standard deviation of downside deviations. In contrast to the IR metric, the Sortino Ratio excludes positive returns from the calculation of volatility, thus not penalizing positive returns in risk assessment. It recognizes the fact that positive and negative returns should not be treated in the same way while calculating the risk-adjusted return metrics (Kidd, 2012).

$$SR = \frac{ARC}{ASD^{-}} \qquad [52]$$

where:

$ASD^{-}$ – annualized standard deviation of downside deviations (negative returns).

## SECTION III

## Data

### 3.1. Data description

In order to perform an empirical assessment of the methodologies described earlier, a set of four worldwide market indices was chosen. These are: S&P 500 (USA), WIG20 (Poland), DAX (Germany) and Nikkei 225 (Japan). The same choice of financial assets was made in Gajda and Walasek (2020). However, in this research, the latest available data is selected, and novel methods of data differentiation are evaluated. The price series data covers the 10 years lasting from 01.01.2014 to 31.12.2023. Depending on the individual asset, this period translates to around 2500 daily observations. Data was downloaded from two sources: Yahoo Finance with the use of Python's *yfinance* library (S&P 500, DAX, and Nikkei 225) and the Stooq portal (WIG20). Figure 3 and Figure 4 present, respectively, the price levels and the corresponding logarithmic returns over the entire period.

**Figure 3.** Price levels of chosen market indices in years 2014-2023

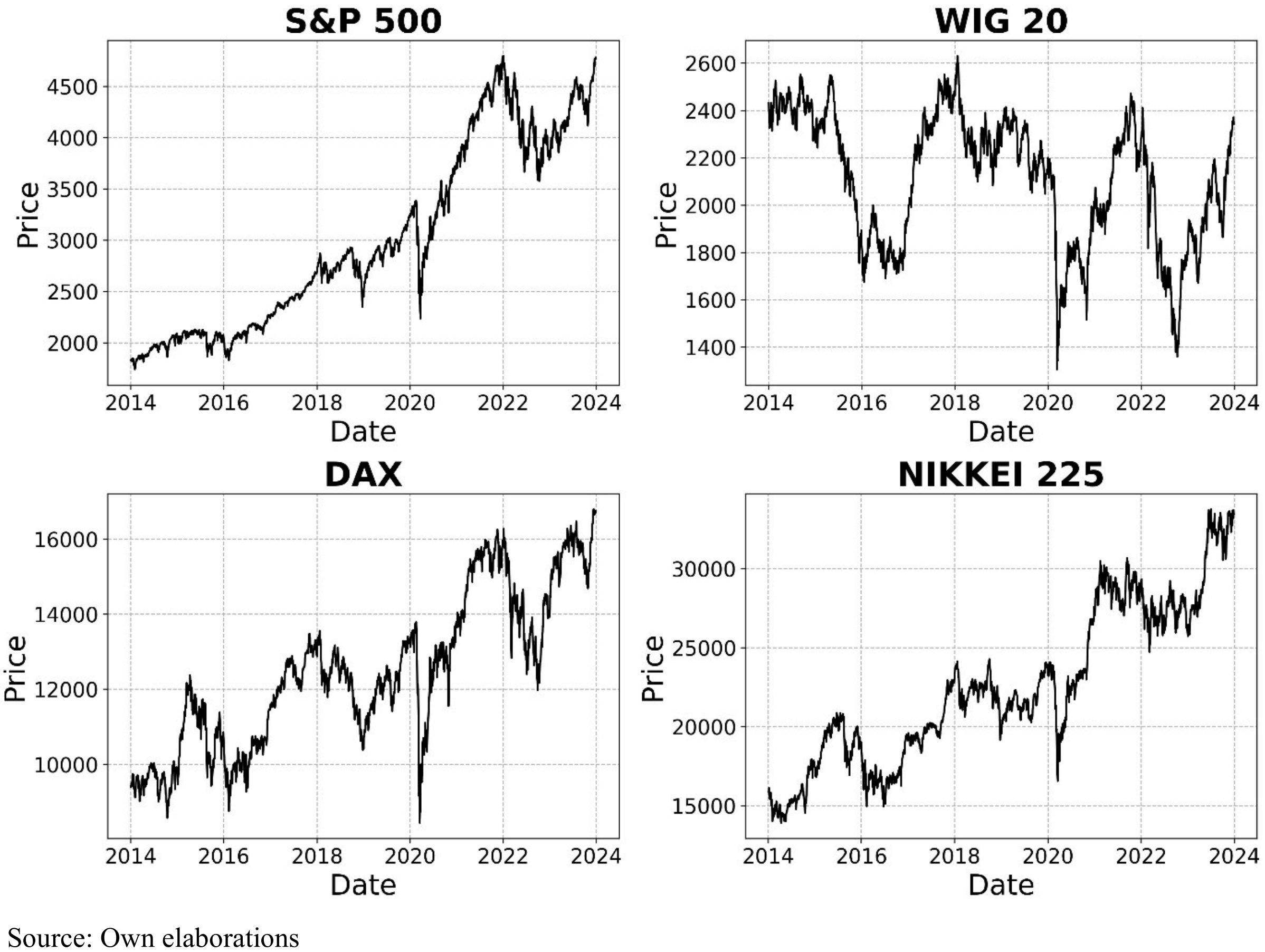

Source: Own elaborations

**Figure 4.** Logarithmic returns of chosen market indices in years 2014-2023

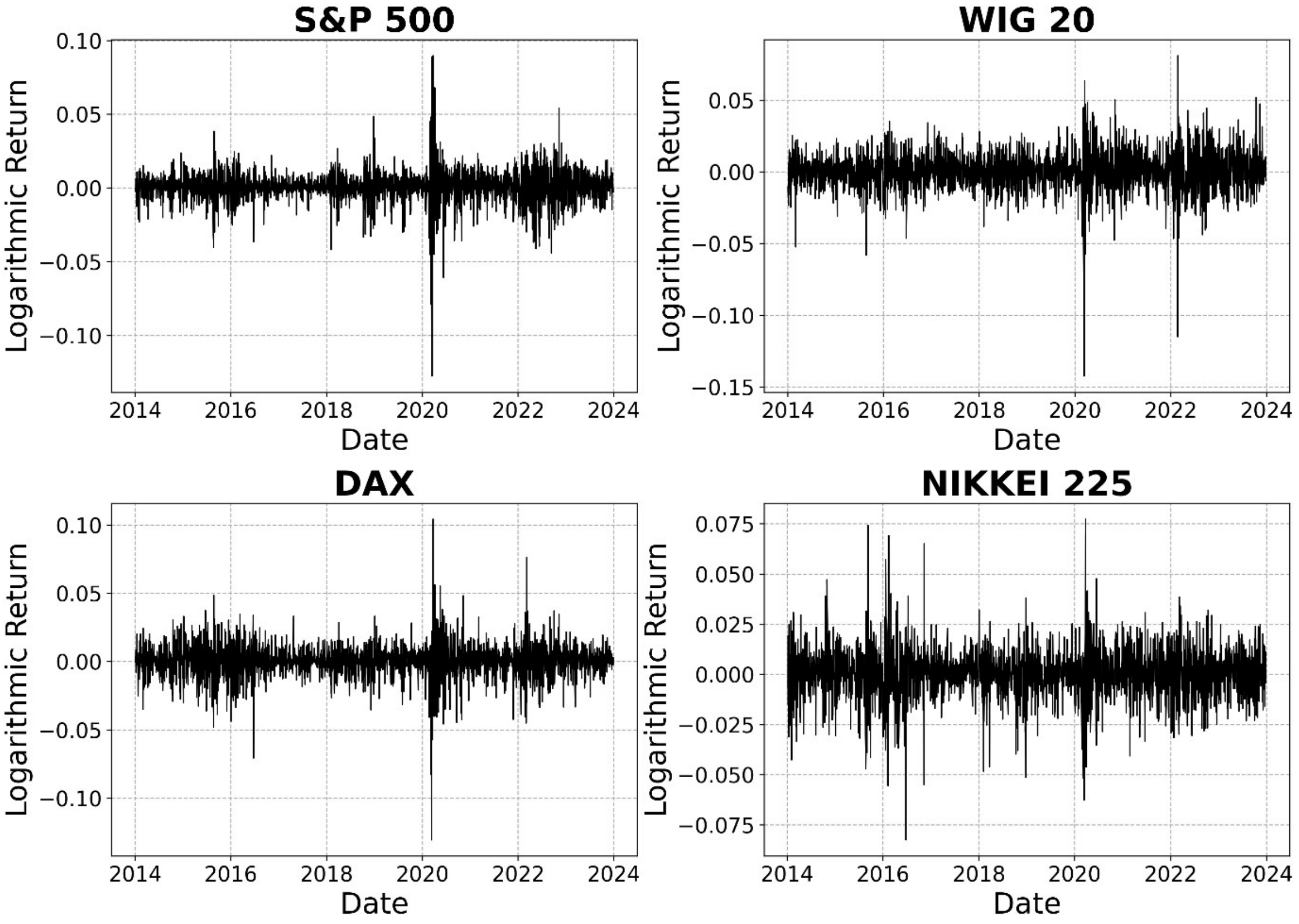

Source: Own elaborations

Based on Figure 3, all market indices, except WIG 20, were characterized by a long-term positive trend with short-term fluctuations. In the case of WIG 20, there is no trend visible in the data, with basically the same value of market index on the first and last day of the chosen period. It is worth noting that all the assets experienced significant and rapid losses in value at the beginning of 2020. The same effect of the worldwide outbreak of the COVID-19 pandemic is represented in Figure 4, where markets exhibited an uncertainty in the form of increased volatility. Moreover, graphically represented logarithmic return series demonstrate a behavior known as volatility clustering, where periods of large changes in prices cluster together and are divided by periods of more stable market conditions. This observation is characteristic for different categories of real-life financial data.

Due to the fact that the main application of the datasets in this research is to serve as an input to the LSTM neural network, the data is divided into three separate groups. The period from 01.01.2014 to 31.12.2020, that is, the first seven years of data, will be utilized as the in-sample training set. On the other hand, the period between 01.01.2021 to 31.12.2023, constituting the last three years of data, will be employed as the out-of-sample testing set.

Moreover, the last two years of training set (period from 01.01.2019 to 31.12.2020) are meant for the validation set and used for the hyperparameter tuning process. So, training and validation sets are intended for network training and finding an optimal set of hyperparameters. In contrast, the testing set is completely unused during this process. Its function is to evaluate the network's performance on the unseen data and draw conclusions concerning the model's potential usability in real-market applications. The described division into training, validation, and testing sets allows for a proper analysis and minimizes the risk of data leakage.

### 3.2. Results of estimations

In this section, the empirical estimation of values of the parameters $d_{De\ Prado}$, $d_{ARFIMA}$ $d_{ARTFIMA}$ and $\lambda_{ARTFIMA}$ takes place, according to the described earlier methodology. This procedure constitutes a crucial part of this research because it enables obtaining differently differentiated price series which are further used as an input for compared deep neural network models. It is worth noting that all the differentiation techniques are performed using the log-transformed price series data.

#### 3.2.1. Estimation of $d_{De\ Prado}$ parameter

The estimation process begins with finding the values of parameter $d_{De\ Prado}$ for all the assets separately based on De Prado's (2018) fractional differencing procedure. The same approach might also be found in Gajda and Walasek (2020). Recalling the framework, the goal is to find a minimum value of the fractional difference operator for which the analyzed price series is characterized by stationary properties. In order to declare whether the transformed series is already stationary, an augmented Dickey-Fuller test (ADF test) is employed, assuming a 95% confidence level. Moreover, to illustrate how closely the fractionally differentiated series resembles the original one, the Pearson coefficient of correlation is calculated for different values of $d$. Estimation of parameter $d_{De\ Prado}$ is conducted on the training part of the data and the obtained optimum value is then applied to the whole dataset. Figures 5-8 illustrate the method of finding an optimum $d_{De\ Prado}$ parameter's value for each market index separately. On the X-axis, increasing values of the fractional difference coefficient are presented. On the other hand, on the Y-axis, two sets of information are placed: the magnitude of correlation between the original series and the fractionally differentiated one (right Y-axis), and the value of the ADF test statistic for the fractionally differentiated series (left Y-axis). Moreover, the horizontal dashed line is positioned at the level of -2.86, corresponding to the critical value of the ADF test at a 95% confidence level.

**Figure 5.** Comparison of ADF test statistic and Pearson coefficient of correlation for various values of fractional differencing parameter implemented for S&P 500 (USA)

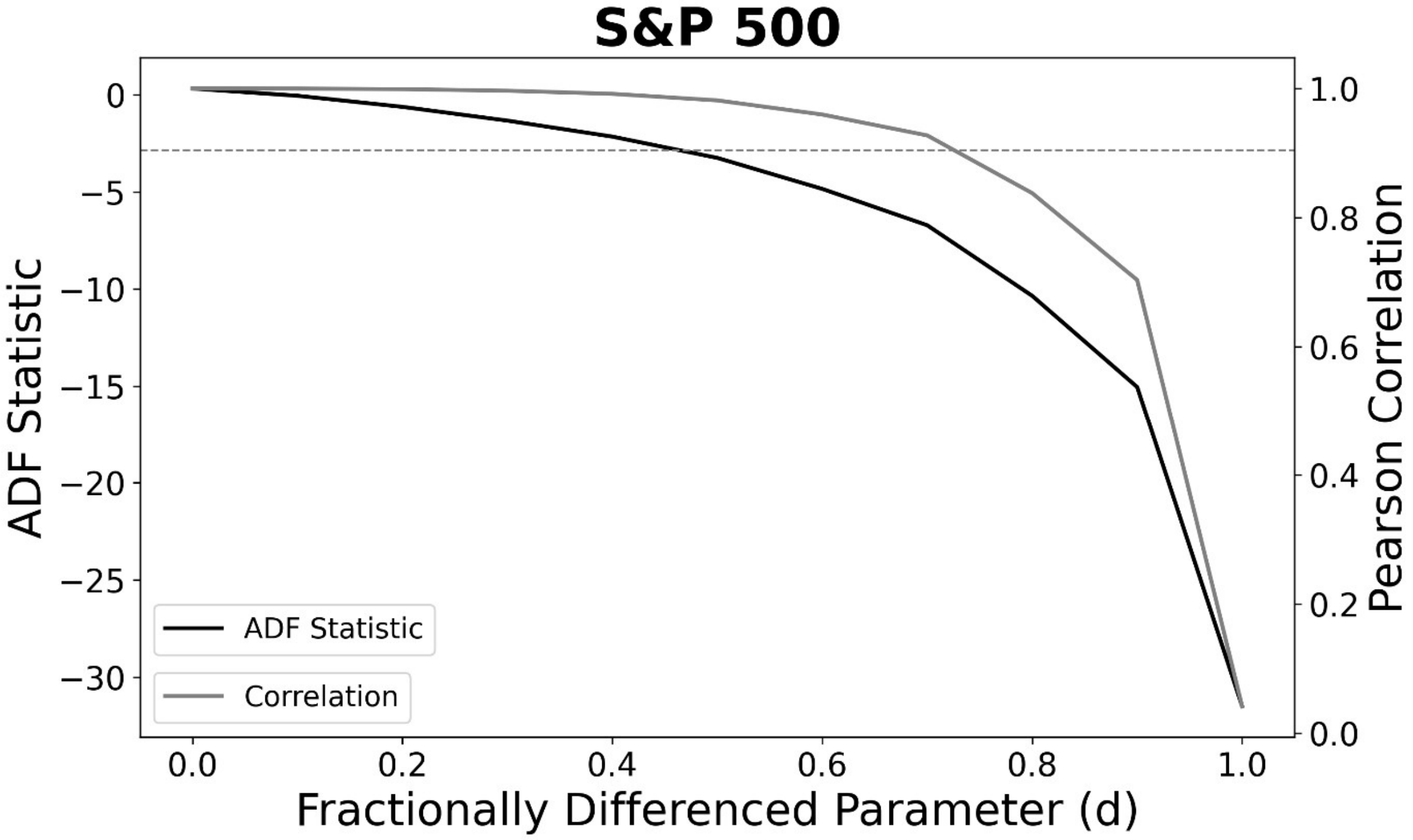

Source: Own elaborations based on De Prado (2018)

**Figure 6.** Comparison of ADF test statistic and Pearson coefficient of correlation for various values of fractional differencing parameter implemented for WIG20 (Poland)

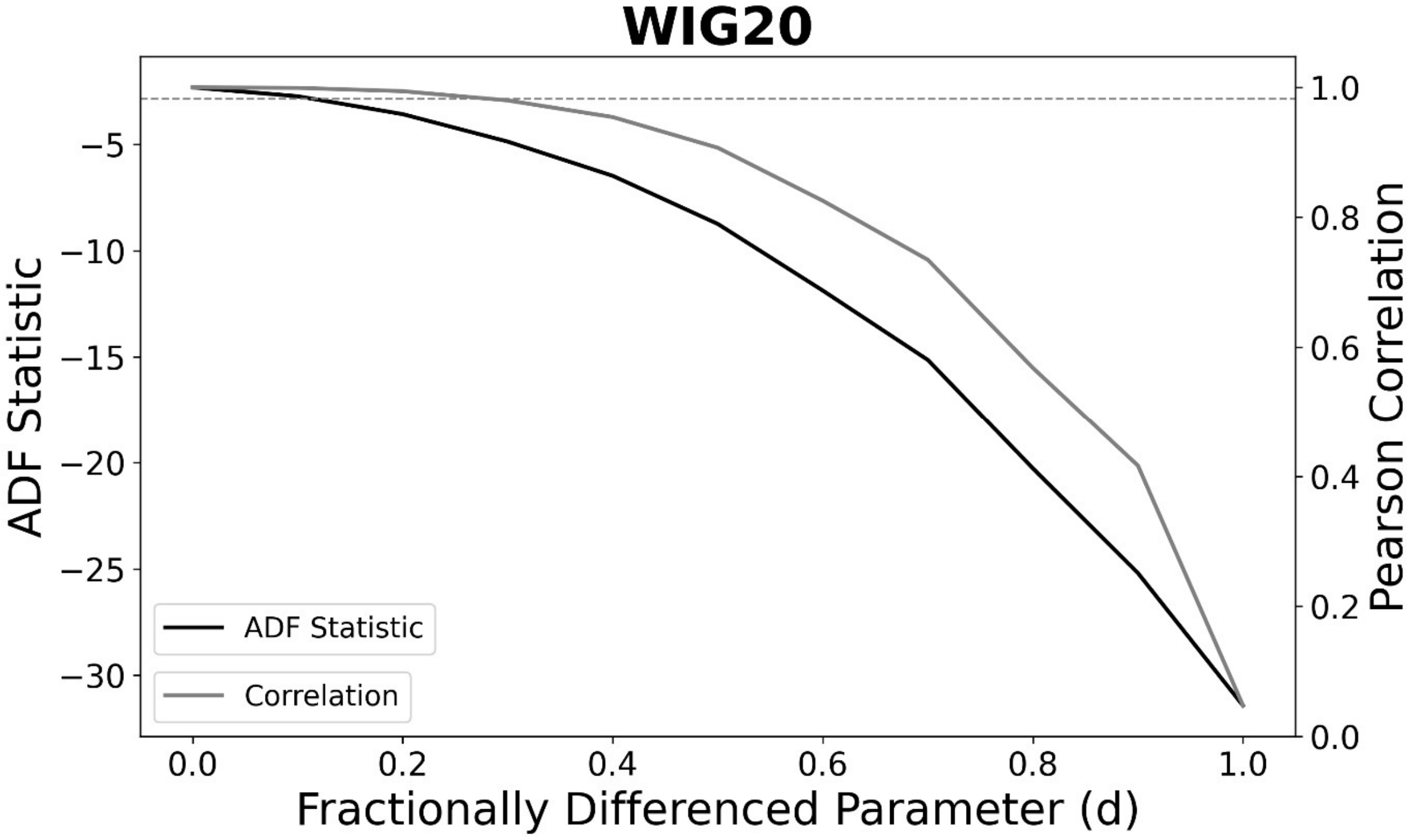

Source: Own elaborations based on De Prado (2018)

**Figure 7.** Comparison of ADF test statistic and Pearson coefficient of correlation for various values of fractional differencing parameter implemented for DAX (Germany)

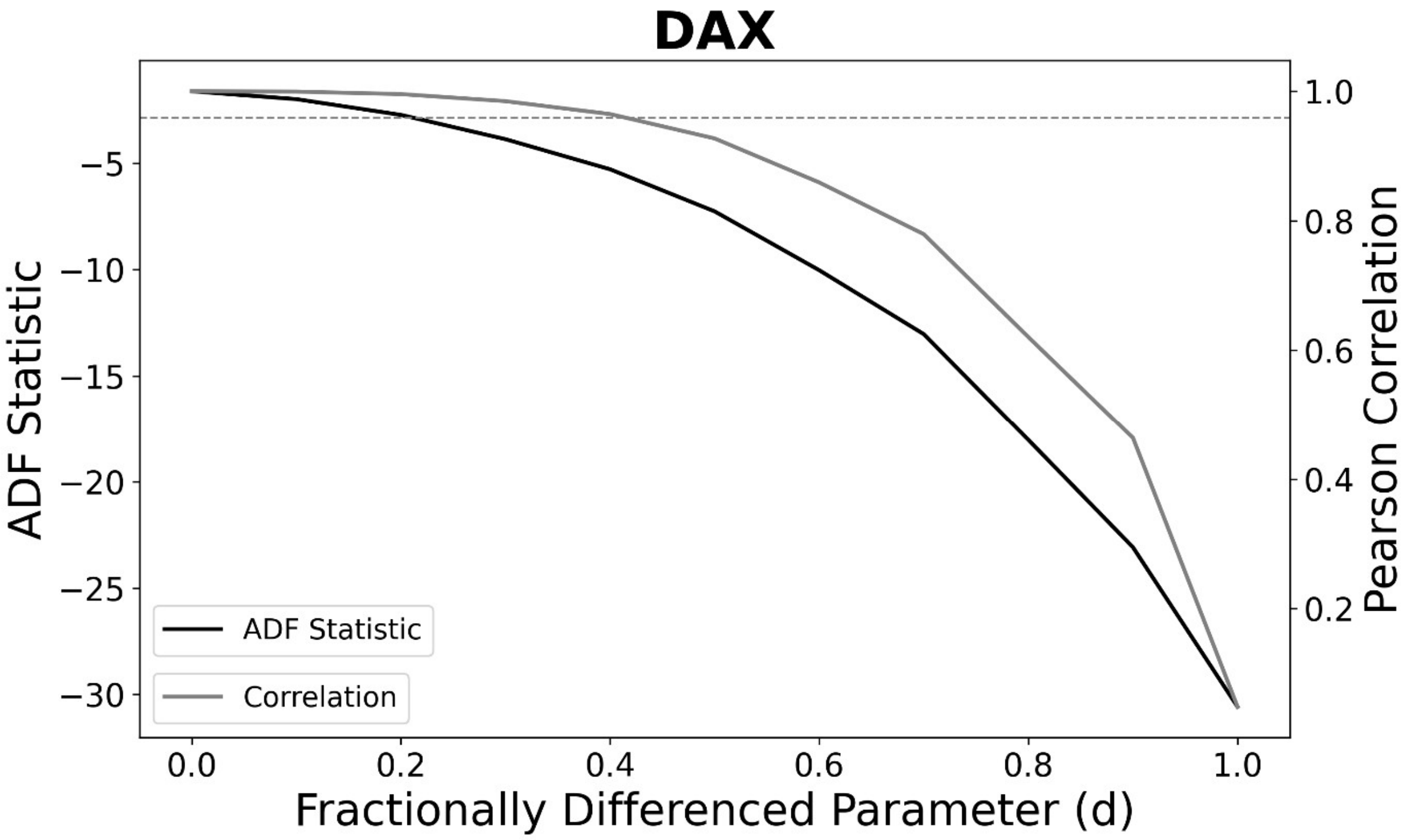

Source: Own elaborations based on De Prado (2018)

**Figure 8.** Comparison of ADF test statistic and Pearson coefficient of correlation for various values of fractional differencing parameter implemented for Nikkei 225 (Japan)

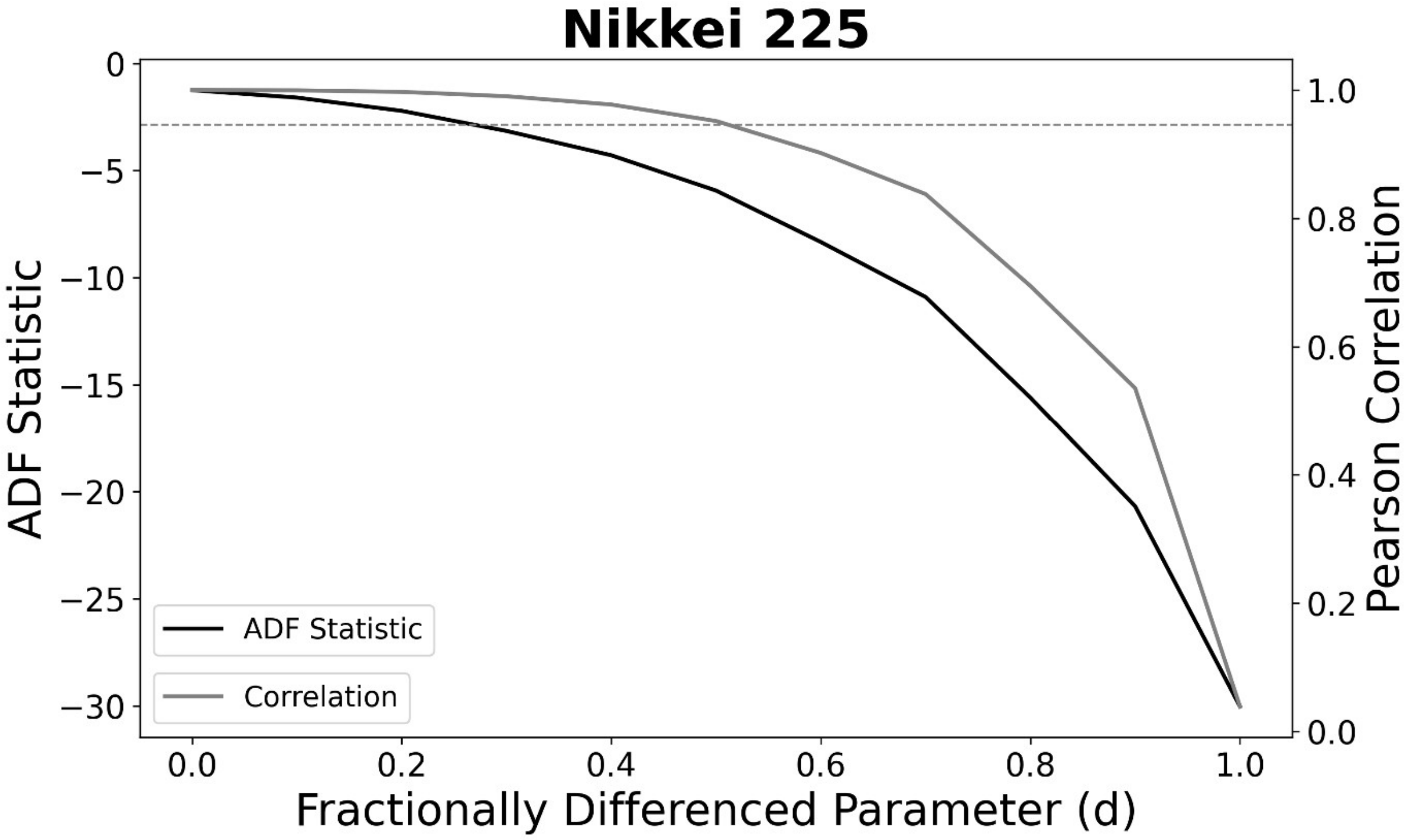

Source: Own elaborations based on De Prado (2018)

Focusing on Figure 5, it presents the process of finding an optimal value of $d_{De\ Prado}$ parameter for S&P 500 market index. It is clearly visible that for values of the $d$ parameter between 0.4 and 0.6, the ADF test statistic crosses the critical value at the 95% confidence level (-2.86). It signifies that, within this range, there is evidence to reject the null hypothesis of non-stationarity. Consequently, after performing De Prado's (2018) fractional differentiation procedure for the $d$ within this interval, the transformed series becomes stationary according to the ADF test. At the same time, the coefficient of correlation between the original and transformed series is still significantly high, having a value of approximately 0.99. For $d = 1$, that is, in the case of full-order differentiation, the correlation coefficient is equal to 0.04. That is a good example of a situation where overdifferencing of the time series removes an excessive amount of its internal memory. The figures of the other assets demonstrate similar conclusions: 1) the optimal value of parameter $d$ is significantly smaller than 1; 2) the coefficient of correlation is still very high in the case of fractional differencing; and 3) full order differentiation tends to cut a substantial amount of memory. The only difference is that the estimated values of the $d_{De\ Prado}$ parameter vary across assets. Their exact levels are presented in Table 1 for each market index separately.

**Table 1.** Estimated values of $d_{De\ Prado}$ parameter for different assets

| | **S&P 500** | **WIG20** | **DAX** | **Nikkei 225** |
|---|---|---|---|---|
| **Estimated $d_{De\ Prado}$** | 0.46 | 0.12 | 0.22 | 0.28 |

Source: Own elaborations

The last phase of performing fractional differencing is to apply estimated values of the $d_{De\ Prado}$ parameter to the analyzed series to obtain their fractionally differenced form. Figure 9 demonstrates the values and dynamics of the fractionally differenced series using the estimated $d_{De\ Prado}$ parameter for each market index separately. Additionally, the log-transformed original series is plotted in order to compare both series. Obtained series $\Delta^{d_{De\ Prado}} X_t$ will be used in the further part of the research as one of the inputs for constructed LSTM models in order to assess an applicability of this type of data transformation for financial assets predictions.

**Figure 9.** Comparison of original log-transformed series (black) and fractionally differenced series (grey) for different assets in years 2014-2023

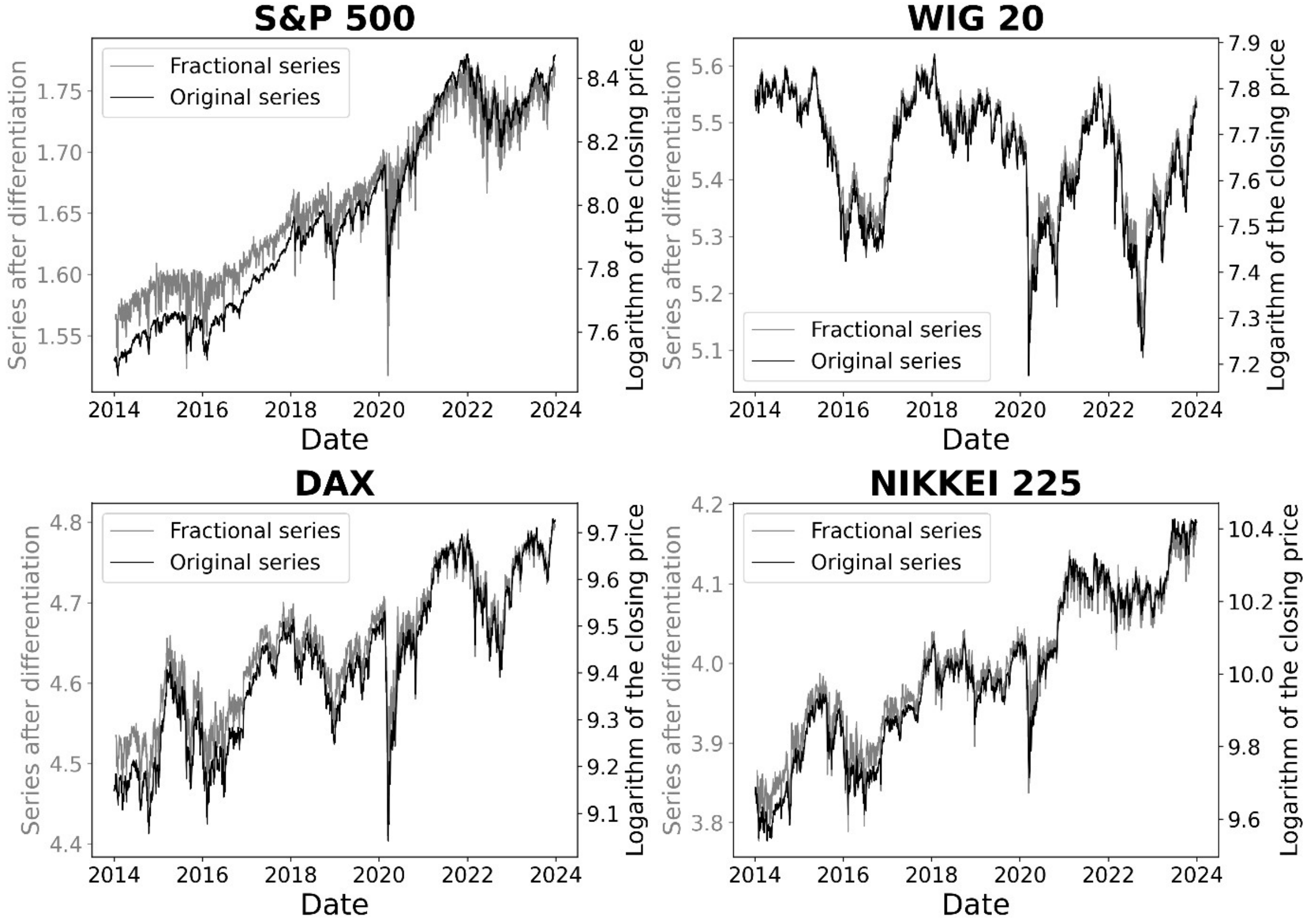

Source: Own elaborations

### 3.2.2. Estimation of $d_{ARFIMA}$, $d_{ARTFIMA}$ and $\lambda_{ARTFIMA}$ parameters

The next stage of the research consists of estimating the $d_{ARFIMA}$, $d_{ARTFIMA}$ and $\lambda_{ARTFIMA}$ parameters. The obtained values are then applied to the price series in order to compute the fractional and tempered fractional differentiated series. The estimation process is performed in the *R* programming language with the use of *artfima* package which facilitates fitting the ARTFIMA and ARFIMA models (Sabzikar et al., 2019). Parameter $d_{ARFIMA}$ is estimated by fitting the data to ARFIMA$(0, d, 0)$ model, while $d_{ARTFIMA}$ and $\lambda_{ARTFIMA}$ parameters are derived by ARTFIMA$(0, d, \lambda, 0)$ model. Tables 2 and 3 present the estimated parameter values for each asset separately, and for the ARFIMA and ARTFIMA models respectively.

**Table 2.** Estimated values of $d_{ARFIMA}$ parameter for different assets

| | **S&P 500** | **WIG20** | **DAX** | **Nikkei 225** |
|---|---|---|---|---|
| **Estimated $d_{ARFIMA}$** | 0.4892 | 0.4892 | 0.4895 | 0.4891 |

Source: Own elaborations

**Figure 10.** Comparison of original log-transformed series (black), fractionally differenced series (grey) and fractionally differenced series based on parameter estimated by ARFIMA model (blue) for different assets in years 2014-2023

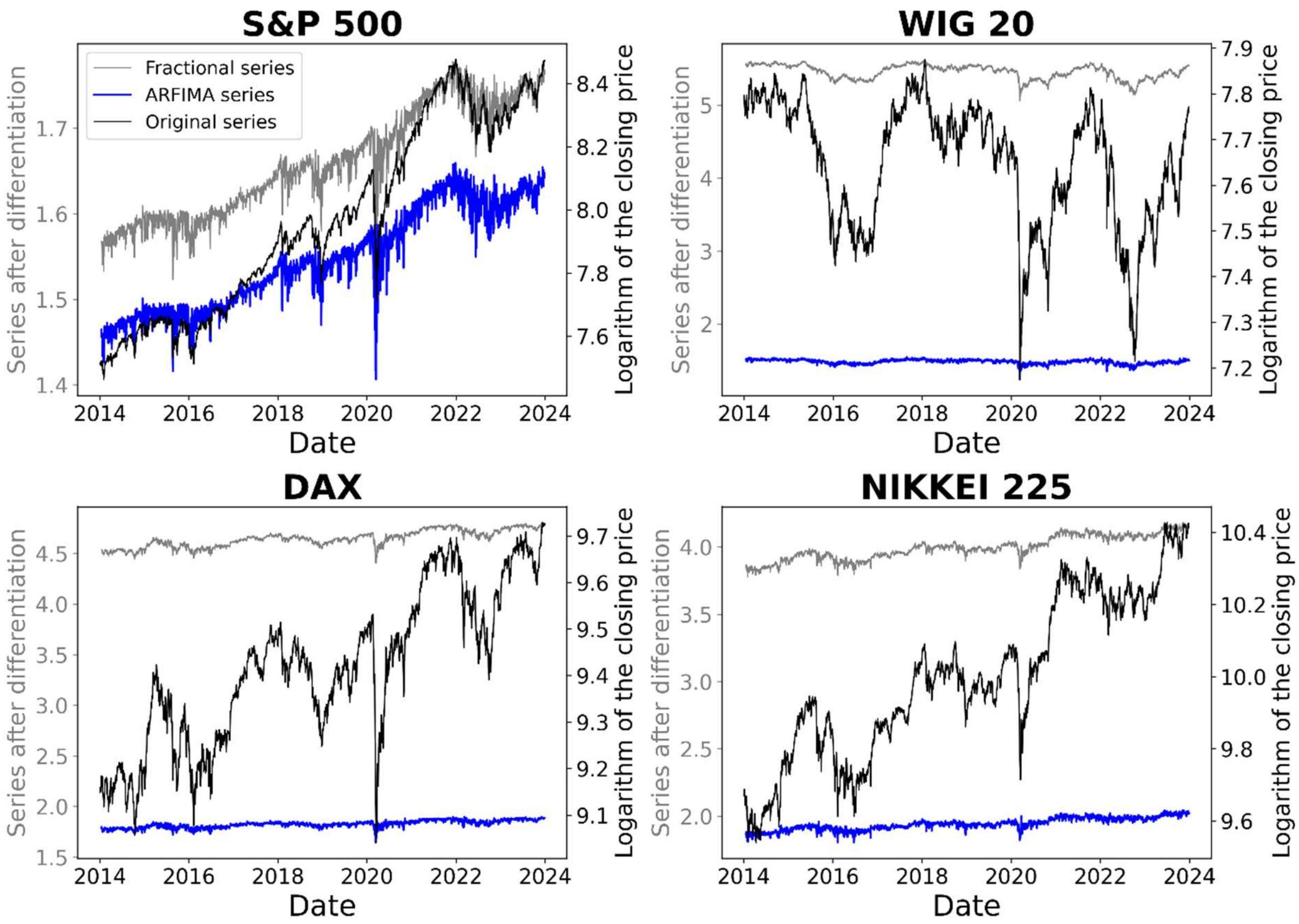

Source: Own elaborations

Analyzing Table 2, the estimated values of the $d_{ARFIMA}$ parameter appear to be relatively stable and close to 0.5. It is worth mentioning that they are all greater than their equivalent $d_{De\ Prado}$ parameters. On the other hand, the information presented in Table 3 indicates that the estimated values of the $d_{ARTFIMA}$ parameter are also consistent and approximately equal to 1. Estimations of the $\lambda_{ARTFIMA}$ parameter are characterized by small but more diverse values – the $\lambda_{ARTFIMA}$ parameter for WIG20 is approximately 20 times bigger than in the case of S&P 500 index.

**Table 3.** Estimated values of $d_{ARTFIMA}$ and $\lambda_{ARTFIMA}$ parameters for different assets

| | S&P 500 | WIG20 | DAX | Nikkei 225 |
|---|---|---|---|---|
| **Estimated $d_{ARTFIMA}$** | 0.9895 | 1.0187 | 1.0050 | 0.9953 |
| **Estimated $\lambda_{ARTFIMA}$** | 0.0003 | 0.0059 | 0.0030 | 0.0016 |

Source: Own elaborations

**Figure 11.** Comparison of logarithmic returns series (black) and tempered-fractionally differenced series (blue) based on parameters estimated by ARTFIMA model for different assets in years 2014-2023

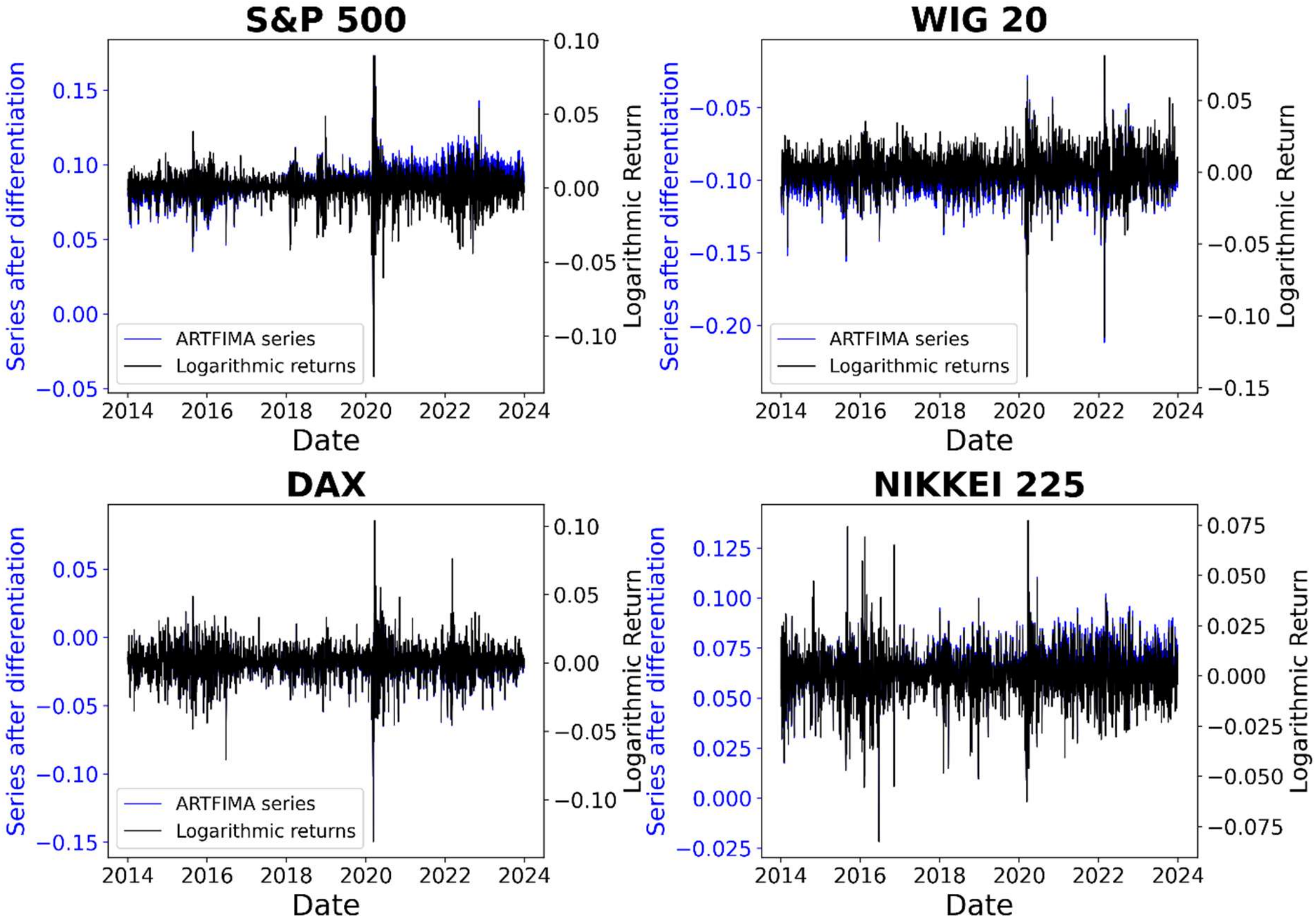

Source: Own elaborations

After acquiring the estimated parameter values from the ARFIMA$(0, d, 0)$ and ARTFIMA$(0, d, \lambda, 0)$ models, the next phase is to utilize them in De Prado's (2018) fractional differencing method and its extended version, which acknowledges an additional tempering parameter. Figure 10 illustrates a comparison of original log-transformed series, fractionally differenced series with $d_{De\ Prado}$ parameter, and fractionally differenced series based on the parameter estimated by the ARFIMA model for different assets. The fractionally differenced series using the $d_{ARFIMA}$ parameter still bears a resemblance to the other two series, however, it appears significantly more flattened. Depending on the asset, the coefficient of correlation with the original series ranges from 0.91 to 0.98. These values are still significantly greater than in the case of full-order differentiation and only slightly smaller in comparison to the fractional series obtained with $d_{De\ Prado}$ parameter. Figure 11 presents a comparison of logarithmic returns and fractionally differenced series with tempering based on the parameters estimated by the ARTFIMA model. Both series closely resemble each other, suggesting that the parameters estimated in Table 3 modify the log-transformed prices into a series that is highly similar to the logarithmic returns. This may indicate that an excessive amount of memory has been removed in the process.

The estimated parameters, when applied to the price series, result in the fractionally differentiated series $\Delta^{d_{ARFIMA}} X_t$ and the fractionally differentiated series with tempering $\Delta^{d_{ARTFIMA}, \lambda_{ARTFIMA}} X_t$. Along with the logarithmic returns $\Delta X_t$ and the $\Delta^{d_{De\ Prado}} X_t$ series, these will be used in the next stage of the research, which involves generating predictions using LSTM models.

# SECTION IV

## Results

### 4.1. An overview of the research procedure

Before discussing the obtained results and due to the significant variety and complexity of the applied techniques, this section briefly summarizes the individual stages of the research procedure. Most of the procedures were carried out using the *Python* programming language. An exception was made for estimating the parameters of the ARFIMA and ARTFIMA models, which were implemented using the *R* statistical software. To obtain the final results, several stages were necessary, and they were performed in the following order:

1. Downloading and preprocessing of the data containing prices of four market indices: S&P 500, WIG20, DAX, and Nikkei 225 for the period from 01.01.2014 to 31.12.2023.
2. Dividing the data into training (first 7 years) and testing periods (last 3 years). Moreover, the last 2 years of the training period are used as the validation data.
3. Calculating logarithmic returns – $\Delta X_t$ .
4. Estimation of the $d_{De\ Prado}$, $d_{ARFIMA}$, $d_{ARTFIMA}$ and $\lambda_{ARTFIMA}$ parameters using training part of the data,
5. Applying estimated values of the parameters to the whole data period and computing the corresponding $\Delta^{d_{De\ Prado}} X_t$ , $\Delta^{d_{ARFIMA}} X_t$ , $\Delta^{d_{ARTFIMA}, \lambda_{ARTFIMA}} X_t$ series.
6. Creating a set of independent variables: the lagged value of the series and several technical indicators – $SMA_5$, $SMA_{10}$, $SMA_{20}$, $RSI_9$, $RSI_{14}$, $RSI_{21}$, $Upper\ BBand_{10,2}$, $Lower\ BBand_{10,2}$, and $MACD_{12,26,9}$.
7. Construction of LSTM networks for each of the $\Delta X_t$ , $\Delta^{d_{De\ Prado}} X_t$ , $\Delta^{d_{ARFIMA}} X_t$ , $\Delta^{d_{ARTFIMA}, \lambda_{ARTFIMA}} X_t$ series and assets independently.
8. Training of the networks and hyperparameter tuning process using the in-sample part of the data. The following set of hyperparameters was optimized: batch size, epochs number, number of layers, number of cells in each layer, dropout rate, recurrent dropout rate, regularization rate, and learning rate.
9. Applying the trained networks with an optimized set of hyperparameters to out-of-sample testing data in order to generate one-day-ahead predictions on the unseen data.
10. Transformation of the obtained predictions with the use of inverse operations to the applied differentiation techniques in order to enable consistent comparison across all forecast outputs.

11. Comparison of the forecast performance across different differencing methods using RMSE, MAE, and MAPE metrics.
12. Utilizing the generated predictions to construct $Long - Short$ and $Long\ Only$ trading signals.
13. Backtesting the performance of each strategy using the testing set, including transactions costs.
14. Constructing and backtesting a trading strategy that treats all the assets as a single portfolio.
15. Comparing the performance of the strategies using trading performance metrics.

### 4.2. Results

After the LSTM networks training procedure and finding optimal sets of hyperparameters for each asset and each data differentiation technique, the next step involves generating predictions on the out-of-sample testing dataset. To enable comparison of the obtained forecasts, inverse operations corresponding to the applied differentiation techniques are performed. Consequently, the transformed predictions are brought back to the original price level. Table 4 presents the attained forecast accuracy based on selected evaluation metrics for each of the market indices separately. Logarithmic returns series is denoted as $\Delta X_t$, fractional differentiation series with $d_{De\ Prado}$ parameter as $\Delta^{d_{De\ Prado}} X_t$, fractional differentiation series with $d_{ARFIMA}$ parameter as $\Delta^{d_{ARFIMA}} X_t$, and fractional differentiation series with tempering with $d_{ARTFIMA}$ and $\lambda_{ARTFIMA}$ parameters as $\Delta^{d_{ARTFIMA}, \lambda_{ARTFIMA}} X_t$.

The presented results strongly indicate that, for each market index, one of the methods of fractional differentiation is characterized by the most accurate one-day-ahead predictions, as evidenced by the lowest values of RMSE, MAE and MAPE metrics. In the case of S&P 500 and Nikkei 225 stock indices, both $\Delta^{d_{De\ Prado}} X_t$ and $\Delta^{d_{ARFIMA}} X_t$ achieved comparable results with fractionally differentiated series by De Prado’s (2018) method being slightly more efficient. For WIG20 index, $\Delta^{d_{ARFIMA}} X_t$ appeared to be superior to all other methods. On the other hand, forecasts generated for $\Delta^{d_{De\ Prado}} X_t$ were the least accurate among all techniques of data differentiation. A possible explanation of this observation is that the estimated $d_{De\ Prado}$ parameter for WIG20 was significantly smaller than for the other assets. Although this value was sufficient to obtain stationary time-series according to the ADF test, the applied order of differentiation might have been too small to make LSTM model handle it effectively. For the $d_{ARFIMA}$ parameter, which is greater than the value of $d_{De\ Prado}$ but still strongly smaller than 1, LSTM network generated the most accurate predictions for all three evaluation metrics. Finally, in forecasting unseen values of DAX index, model employing $\Delta^{d_{De\ Prado}}X_t$ series produced the most effective predictions. In summary, for all four market indices the most accurate method of one-day-ahead forecasting was either the fractionally differentiated series with the $d_{De\ Prado}$ parameter – $\Delta^{d_{De\ Prado}}X_t$ or the fractionally differentiated series with the $d_{ARFIMA}$ parameter – $\Delta^{d_{ARFIMA}}X_t$. For three out of the four assets, excluding the WIG20 stock index, these two methods emerged as the top performers.

**Table 4.** Comparison of LSTM network's results on the testing set depending on different techniques of data differentiation

| Market index | Method | RMSE | MAE | MAPE |
|---|---|---|---|---|
| S&P 500 | $\Delta X_t$ | *65.30* | *49.85* | *1.19%* |
| | $\Delta^{d_{De\ Prado}}X_t$ | ***49.81*** | *39.20* | *0.94%* |
| | $\Delta^{d_{ARFIMA}}X_t$ | *49.94* | ***38.58*** | ***0.93%*** |
| | $\Delta^{d_{ARTFIMA}, \lambda_{ARTFIMA}}X_t$ | *59.17* | *43.91* | *1.06%* |
| WIG20 | $\Delta X_t$ | *33.30* | *24.32* | *1.25%* |
| | $\Delta^{d_{De\ Prado}}X_t$ | *39.43* | *30.85* | *1.57%* |
| | $\Delta^{d_{ARFIMA}}X_t$ | ***30.53*** | ***22.87*** | ***1.17%*** |
| | $\Delta^{d_{ARTFIMA}, \lambda_{ARTFIMA}}X_t$ | *32.42* | *24.21* | *1.25%* |
| DAX | $\Delta X_t$ | *185.91* | *137.47* | *0.94%* |
| | $\Delta^{d_{De\ Prado}}X_t$ | ***162.39*** | ***118.89*** | ***0.81%*** |
| | $\Delta^{d_{ARFIMA}}X_t$ | *175.11* | *133.20* | *0.90%* |
| | $\Delta^{d_{ARTFIMA}, \lambda_{ARTFIMA}}X_t$ | *212.71* | *164.89* | *1.11%* |
| Nikkei 225 | $\Delta X_t$ | *375.75* | *290.99* | *1.01%* |
| | $\Delta^{d_{De\ Prado}}X_t$ | ***344.09*** | ***268.10*** | ***0.93%*** |
| | $\Delta^{d_{ARFIMA}}X_t$ | *344.46* | *271.31* | *0.94%* |
| | $\Delta^{d_{ARTFIMA}, \lambda_{ARTFIMA}}X_t$ | *397.92* | *312.57* | *1.09%* |

Source: Own elaborations

With regard to the logarithmic returns series – $\Delta X_t$ and the fractionally differentiated series with tempering – $\Delta^{d_{ARTFIMA}, \lambda_{ARTFIMA}}X_t$, these two methods are characterized by the lower quality forecasting performance. For almost all assets, the values of evaluation metrics indicate that they are inferior to fractional differencing methods. Additionally, there is no clear pattern among them which technique is better as a data preprocessing method that generates transformed input for forecasting models. The obtained results might provide an evidence for De Prado's (2018) argument that overdifferencing of time-series data removes an excessive amount of memory which could make forecasting more complicated and less accurate. However, the computed results also suggest that the smallest value of differencing parameter that makes series stationary ($d_{De\ Prado}$) does not always lead to the most accurate predictive model. In some cases, an alternative real value of differencing operator (such as $d_{ARFIMA}$) that remains significantly below one might result in superior predictions produced by the LSTM network based on correspondingly fractionally differentiated series.

### 4.3. Backtesting of trading strategies

In this section, the obtained predictions are transformed into trading signals for the testing period, in line with $Long - Short$ and $Long\ Only$ strategies. The evaluation of the generated equity lines is based on previously described trading performance metrics, with particular emphasis on risk-adjusted return metrics such as the Information Ratio and Sortino Ratio. Firstly, the performance of strategies for different methods of data differencing is compared for each market index separately. For this purpose, the equity lines are visualized, and the corresponding performance metrics are presented. Moreover, the constructed equity lines for different methodologies are evaluated against the benchmark, defined as a simple *Buy & Hold* strategy applied to each asset. Subsequently, all assets are treated as a single equally weighted portfolio, and the effectiveness of this approach is summarized analogously.

#### 4.3.1. Individual stock indices

Figures 12 – 19 (placed in the Annex) illustrate equity lines for all the assets and $Long - Short$ / $Long\ Only$ strategies. Each figure presents a comparison of the benchmark *Buy & Hold* strategy and four strategies generated by inputting LSTM networks with differently differenced price series. Corresponding trading performance metrics for each figure are presented in Tables 5 – 12 (placed in the Annex).

The analysis of the presented results allows for drawing conclusions concerning the practical applicability of the employed methods in a trading environment. Rather than focusing on individual assets, the obtained results will be analyzed collectively. Starting with findings for $Long - Short$ signals, the generated trading strategies rarely managed to beat the benchmark. Regarding the Information Ratio metric, the LSTM models based on the logarithmic returns or fractionally differenced series using $d_{ARFIMA}$ parameter outperformed the market once (WIG 20). In contrast, the other two methods of differentiation did not outperform the market in any instance. However, it is worth noting that the method using fractionally differenced series with the $d_{De\ Prado}$ parameter was twice (DAX and Nikkei 225), based on the IR metric, the best performing one (excluding $Buy\,\&\,Hold$ scenario). With respect to the Sortino Ratio metric, the compared strategies outperformed the benchmark following number of times: $\Delta X_t$ – 2 (S&P500 and WIG20), $\Delta^{d_{De\ Prado}} X_t$ – 0 (although it again yielded the best results for DAX and Nikkei 225), $\Delta^{d_{ARFIMA}} X_t$ – 2 (S&P500 and WIG20), $\Delta^{d_{ARTFIMA}, \lambda_{ARTFIMA}} X_t$ – 0. Based on risk-adjusted return metrics for $Long - Short$ signals, the most effective differentiation method appears to be fractional differentiation with the $d_{ARFIMA}$ parameter, followed by logarithmic returns and fractional differentiation with the $d_{De\ Prado}$ parameter.

Continuing with the $Long\ Only$ signals, fractional differentiation with the $d_{ARFIMA}$ parameter once again proved to be the most efficient method, managing to outperform the benchmark across all assets in terms of the Information Ratio metric. Logarithmic returns (S&P 500 and WIG20) and fractional differentiation with the $d_{De\ Prado}$ parameter (S&P 500 and DAX) achieved it twice. Fractional differentiation with tempering accomplished that only once for the WIG20 index. In the case of the Sortino Ratio metric, the obtained results and conclusions are similar. It may be concluded that, based on the achieved results, the predictions generated by the LSTM networks appeared to be more effective when transformed into $Long\ Only$ trading signals compared to $Long - Short$ signals.

### 4.3.2. Portfolio of the assets

In order to aggregate the results discussed above for the individual stock indices, this section considers all assets as a single portfolio. Financial asset allocation and portfolio optimization constitute another challenging and widely studied topic in the finance literature (Mirete-Ferrer et al, 2022; Gunjan and Bhattacharyya, 2023). Given its relative simplicity, this study adopts an equally weighted portfolio approach to asset management. For each index, an equal amount of capital is allocated, which in an equally weighted portfolio constructed of four assets is equal to 25%. This approach provides a more robust performance assessment of different methods of data differentiation by slightly reducing asset-specific fluctuations. It is worth noting that the LSTM model finds wide application in the field of portfolio management research (Malandri et al., 2018; Obeidat et al., 2018; Yun et al., 2020; Zhang et al., 2020). Figures 20 – 21 illustrate the obtained equity lines for $Long - Short$ and $Long\ Only$ strategies, respectively. Corresponding trading performance metrics are placed in Tables 13 – 14.

**Figure 20.** Equity lines for $Long - Short$ strategy, depending on different techniques of data differentiation for the portfolio of assets

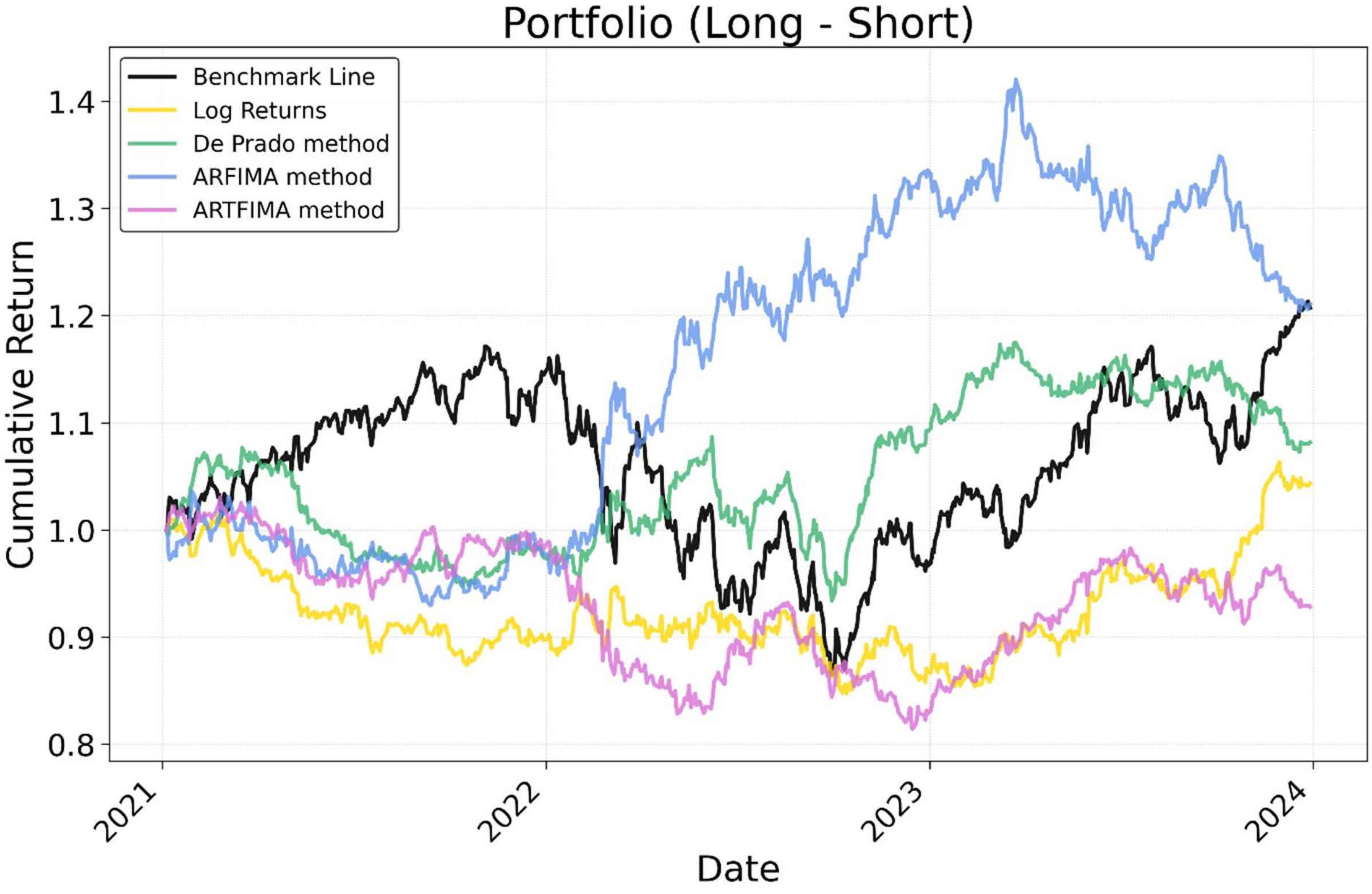

*Equity lines represent a trading performance for different techniques of data differentiation in period from 01.01.2021 to 31.12.2023. Transaction costs of 0.005% are included (Kryńska and Ślepaczuk, 2022).*

Source: Own elaborations

**Table 13.** Trading performance metrics for $Long - Short$ strategy, depending on different techniques of data differentiation for the portfolio of assets

| Indicator | $\Delta X_t$ | $\Delta^{d_{De\ Prado}} X_t$ | $\Delta^{d_{ARFIMA}} X_t$ | $\Delta^{d_{ARTFIMA}, \lambda_{ARTFIMA}} X_t$ | Buy & Hold |
|---|---|---|---|---|---|
| **ARC** | *1.43* | *2.65* | ***6.58*** | *-2.45* | *6.49* |
| **ASD** | *9.79* | ***9.20*** | *11.45* | *9.90* | *13.59* |
| **MD** | *16.47* | ***14.05*** | *15.23* | *20.95* | *26.15* |
| **IR** | *0.15* | *0.29* | ***0.57*** | *-0.25* | *0.48* |
| **SR** | *0.24* | *0.49* | ***1.01*** | *-0.38* | *0.79* |

Source: Own elaborations

**Figure 21.** Equity lines for $Long\ Only$ strategy, depending on different techniques of data differentiation for the portfolio of assets

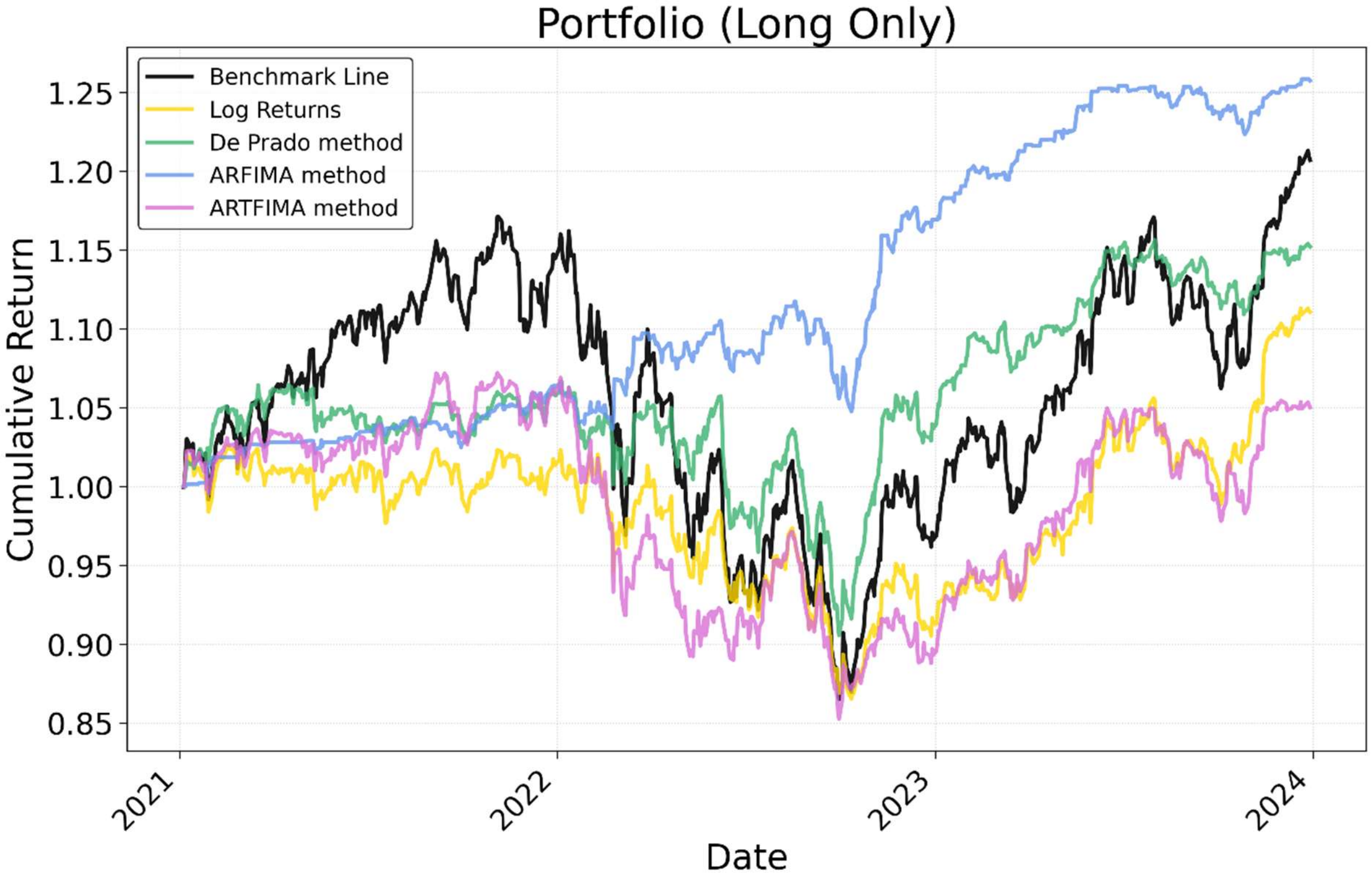

*Equity lines represent a trading performance for different techniques of data differentiation in period from 01.01.2021 to 31.12.2023. Transaction costs of 0.005% are included (Kryńska and Ślepaczuk, 2022).*

Source: Own elaborations

**Table 14.** Trading performance metrics for $Long\ Only$ strategy, depending on different techniques of data differentiation for the portfolio of assets

| Indicator | $\Delta X_t$ | $\Delta^{d_{De\ Prado}} X_t$ | $\Delta^{d_{ARFIMA}} X_t$ | $\Delta^{d_{ARTFIMA},\lambda_{ARTFIMA}} X_t$ | Buy & Hold |
|---|---|---|---|---|---|
| **ARC** | *3.56* | *4.84* | ***7.94*** | *1.65* | *6.49* |
| **ASD** | *9.50* | *8.34* | ***4.86*** | *10.10* | *13.59* |
| **MD** | *15.66* | *15.12* | ***6.23*** | *20.49* | *26.15* |
| **IR** | *0.38* | *0.58* | ***1.63*** | *0.16* | *0.48* |
| **SR** | *0.62* | *0.92* | ***2.99*** | *0.26* | *0.79* |

Source: Own elaborations

The presented results strongly indicate that fractional differentiation techniques outperform not only the rest of the methods, but also the benchmark $Buy \, \& \, Hold$ approach. This observation is true for both $Long - Short$ and $Long \; Only$ signals. As in the case of backtesting based on individual assets, fractional differentiation using the $d_{ARFIMA}$ parameter is characterized by the greatest values for both the Information Ratio and Sortino Ratio metrics. Due to its high annualized rate of return (ARC) and relatively low annualized standard deviation (ASD), the values of risk-adjusted return metrics significantly exceed other methods' results. This is particular evident when analyzing the equity line of the ARFIMA method for $Long \; Only$ signals, which demonstrates steady growth without major drawdowns. The second technique of fractional differentiation, based on the $d_{De \; Prado}$ parameter, also presents relatively good performance, managing to outperform the benchmark approach in terms of the IR and SR metrics for $Long \; Only$ signals. Logarithmic returns and fractionally differenced series with tempering achieved worse results across all trading performance metrics, while the latter was characterized by the least favorable results. These findings provide further support for employing fractional differentiation as a data transformation method in predictive modelling, as opposed to techniques that tend to remove an excessive amount of memory from the time-series.

## CONCLUSIONS

This study aimed to compare different methods of data differentiation in order to find out which one is the most appropriate as an input to the constructed predictive model. The research was inspired by De Prado's (2018) methodology of fractional differentiation, which addresses the issues of overdifferencing and removing an excessive amount of memory from the time-series process. The objective is to identify the minimum real value of the difference operator at which the analyzed series begins to exhibit stationary properties, according to the ADF test. However, this study further extended this approach by including two additional techniques of data differentiation. Firstly, the ARFIMA$(0, d, 0)$ model, introduced by Granger and Joyeux (1980) and Hosking (1981), was employed to estimate an alternative value of the difference operator, which was then used to fractionally differ series. Although the obtained values of $d_{ARFIMA}$ parameter were greater than $d_{De\ Prado}$, the correlation coefficient between the original series and the fractionally differentiated series with parameter from the ARFIMA model was still significantly high. Additionally, the ARTFIMA$(0, d, \lambda, 0)$ model was implemented (Meerschaert et al., 2015), which required an extension of existing methods in order to enable performing fractional differentiation with the tempering parameter. However, after applying fractional differentiation with tempering, the obtained series bore a strong resemblance to the logarithmic returns, suggesting that this novel approach might also represent an excessive memory-removing method. To the author's knowledge, these two extensions of De Prado's (2018) methodology constitute a novelty in the literature and have not been presented before.

Empirical part of this research was based on the four worldwide stock indices: S&P 500 (USA), WIG20 (Poland), DAX (Germany) and Nikkei 225 (Japan). Data covered a 10-year period from the beginning of 2014 until the end of 2023, with the last three years separated as the out-of-sample testing period. In order to generate one-day-ahead predictions, the Long Short-Term Memory (LSTM) recurrent network was employed. The dependent variables included separately differentiated price series ($\Delta X_t$, $\Delta^{d_{De\ Prado}} X_t$, $\Delta^{d_{ARFIMA}} X_t$, and $\Delta^{d_{ARTFIMA}, \lambda_{ARTFIMA}} X_t$) along with a common set of selected technical indicators. Based on the obtained values of performance metrics, the results indicate that the most accurate predictions were generated by LSTM networks which input included the fractionally differentiated price series. This observation supports the findings of Gajda and Walasek (2020). For each asset, the predictive model based on $\Delta^{d_{De\ Prado}} X_t$ or $\Delta^{d_{ARFIMA}} X_t$ was characterized by the lowest values of RMSE, MAE, and MAPE metrics. On the other hand, approaches based on logarithmic returns or fractionally differentiated series with tempering failed to provide reliable forecasts. Obtained findings provide a strong argument against the methods which tend to overdifferentiate the data and thus discard excess memory from the time-series.

The next set of results was obtained after transforming generated predictions into the $Long - Short$ and $Long\ Only$ signals, which were used to backtest them on the testing period of data. The achieved results were evaluated using trading performance metrics, with additional focus put on the risk-adjusted return indicators: Information Ratio and Sortino Ratio. Firstly, the backtesting was performed on individual assets and then on the equally weighted portfolio consisting of all four market indices. Different methods of data differentiation were compared with each other, but also versus the benchmark, which was represented by a simple $Buy\ \&\ Hold$ approach. The acquired equity lines and the corresponding computed metrics strongly suggested that LSTM networks based on fractionally differentiated inputs provide superior performance also in practical applications. The observation is particularly true for the $d_{ARFIMA}$ differencing operator. This method has the most frequently managed to outperform not only other approaches, but also the market. Especially impressive was the performance of this approach for the $Long\ Only$ signals with a portfolio of indices as the underlying asset. The course of the equity line indicated that the strategy was able to beat the market with lower levels of associated risk. The obtained findings provide no grounds to reject the stated research hypotheses regarding the superiority of fractional differencing methods over logarithmic returns in terms of both prediction accuracy and trading performance. Moreover, the best-performing fractional differencing method managed to outperform the market, particularly in the case of the asset portfolio, as indicated by the risk-adjusted return metrics.

The acquired sets of results lead to two main findings that contribute to the research subject. Firstly, the arguments of De Prado (2018) and the results of Gajda and Walasek (2020) were confirmed with the use of new datasets. Fractionally differentiated price series provided a better alternative as inputs to the predictive model, in this case, an LSTM network. This conclusion was confirmed not only by traditional performance metrics but also by trading indicators. So, memory recovering methods proved to be more reliable solutions to employ during forecasting in comparison to logarithmic returns and fractionally differentiated series with tempering. In particular, the poor performance of fractional differencing with tempering leads to the rejection of the hypothesis suggesting its potential advantage over other methods. This technique, similar to logarithmic returns, appears to remove a substantial amount of memory from the time-series process. The second conclusion concerns the choice of the appropriate value of the difference operator during fractional differentiation. The obtained results indicate that the smallest possible value of the differencing operator that achieves stationarity does not always lead to the most reliable predictions. It was demonstrated that a slightly larger $d_{ARFIMA}$ parameter sometimes outperforms the results achieved by applying the $d_{De\ Prado}$ parameter. These findings were especially confirmed in the practical application of predictions to real trading scenarios. Fractionally differentiated series with the $d_{ARFIMA}$ parameter were characterized by significantly lower values of ADF test statistics (indicating stronger evidence in favor of stationarity), while the value of Pearson correlation coefficient with the original series was only slightly lower in comparison to the fractional differentiation with the $d_{De\ Prado}$ parameter.

Summarizing, the empirical analysis conducted in this study provided answers to the previously formulated hypotheses. The findings can be presented as follows:

- *H1:* Supported. Based on the forecasting accuracy metrics, the LSTM models trained on fractionally differentiated data provided more accurate predictions compared to those using logarithmic returns.
- *H2:* Supported. Trading performance metrics further confirmed the superiority of fractionally differentiated data, that is, memory-preserving techniques, as inputs to the predictive model.
- *H3:* Rejected. Fractional differencing with tempering appeared to constitute an approach that tends to remove an excessive amount of memory from the time-series, which negatively affected both forecasting accuracy and trading model performance.
- *H4:* Partially supported. The potential to outperform the benchmark $Buy\ \&\ Hold$ approach depended on the category of trading signals ($Long - Short$ and $Long\ Only$) and the type of underlying asset (individual market indices and the combined portfolio). Fractional differencing methods proved particularly effective for the asset portfolio, as indicated by risk-adjusted rerurn metrics.

At the end, the topic of fractional differentiation is still broad and requires further exploration. Above all, many categories of financial assets should be tested for the most appropriate technique of differentiation. Examples include individual stocks, commodities, currencies, and cryptocurrencies. Also, the analogical study should be performed for market indices of developing countries, as conclusions may differ from the results obtained in this research. Another aspect of possible extensions concerns the method of data splitting. In this study, a simple train–test split was applied. Future research may employ the sliding window approach for time-series cross-validation, which generates multiple train–test splits to make the model learn from the updated data. Another area of future research should be focused on a better understanding of the relationship between the value of the fractional differentiation parameter and predictive model performance. This study revealed that there is no universal method of determining the $d$ parameter value that guarantees the best forecasting and backtesting performance across all assets.

## ANNEXES

**Figure 12.** Equity lines for $Long - Short$ strategy, depending on different techniques of data differentiation for S&P 500

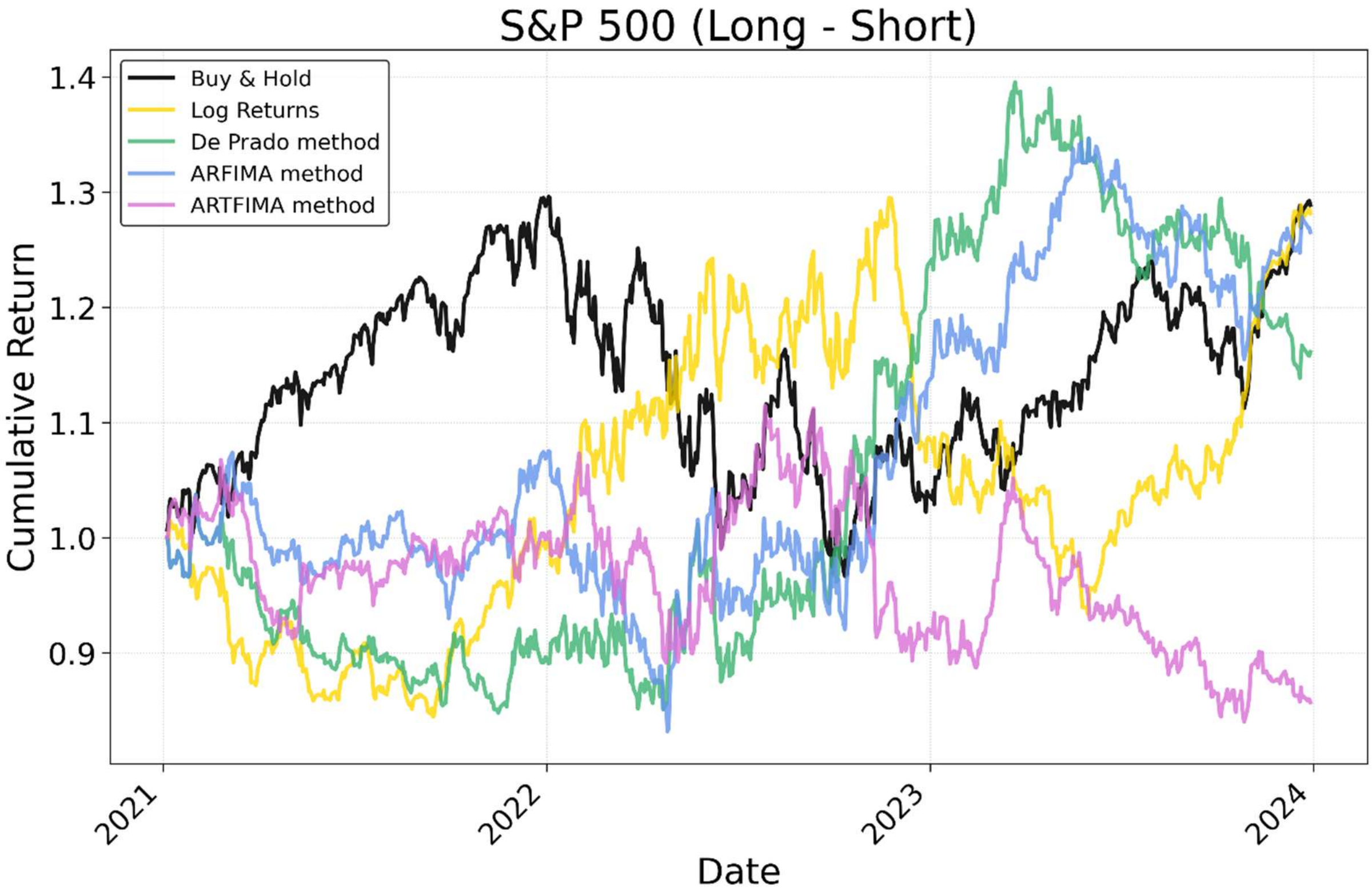

*Equity lines represent a trading performance for different techniques of data differentiation in period from 01.01.2021 to 31.12.2023. Transaction costs of 0.005% are included (Kryńska and Ślepaczuk, 2022).*

Source: Own elaborations

**Table 5.** Trading performance metrics for $Long - Short$ strategy, depending on different techniques of data differentiation for S&P 500

| Indicator | $\Delta X_t$ | $\Delta^{d_{De\ Prado}} X_t$ | $\Delta^{d_{ARFIMA}} X_t$ | $\Delta^{d_{ARTFIMA},\lambda_{ARTFIMA}} X_t$ | Buy & Hold |
|---|---|---|---|---|---|
| **ARC** | *8.63* | *5.11* | *8.16* | *-5.01* | ***8.82*** |
| **ASD** | ***17.48*** | *17.58* | *17.53* | *17.59* | *17.59* |
| **MD** | *27.90* | ***18.42*** | *22.64* | *24.63* | *25.43* |
| **IR** | *0.49* | *0.29* | *0.47* | *-0.29* | ***0.50*** |
| **SR** | ***0.84*** | *0.52* | *0.83* | *-0.45* | *0.83* |

Source: Own elaborations

**Figure 13.** Equity lines for *Long Only* strategy, depending on different techniques of data differentiation for S&P 500

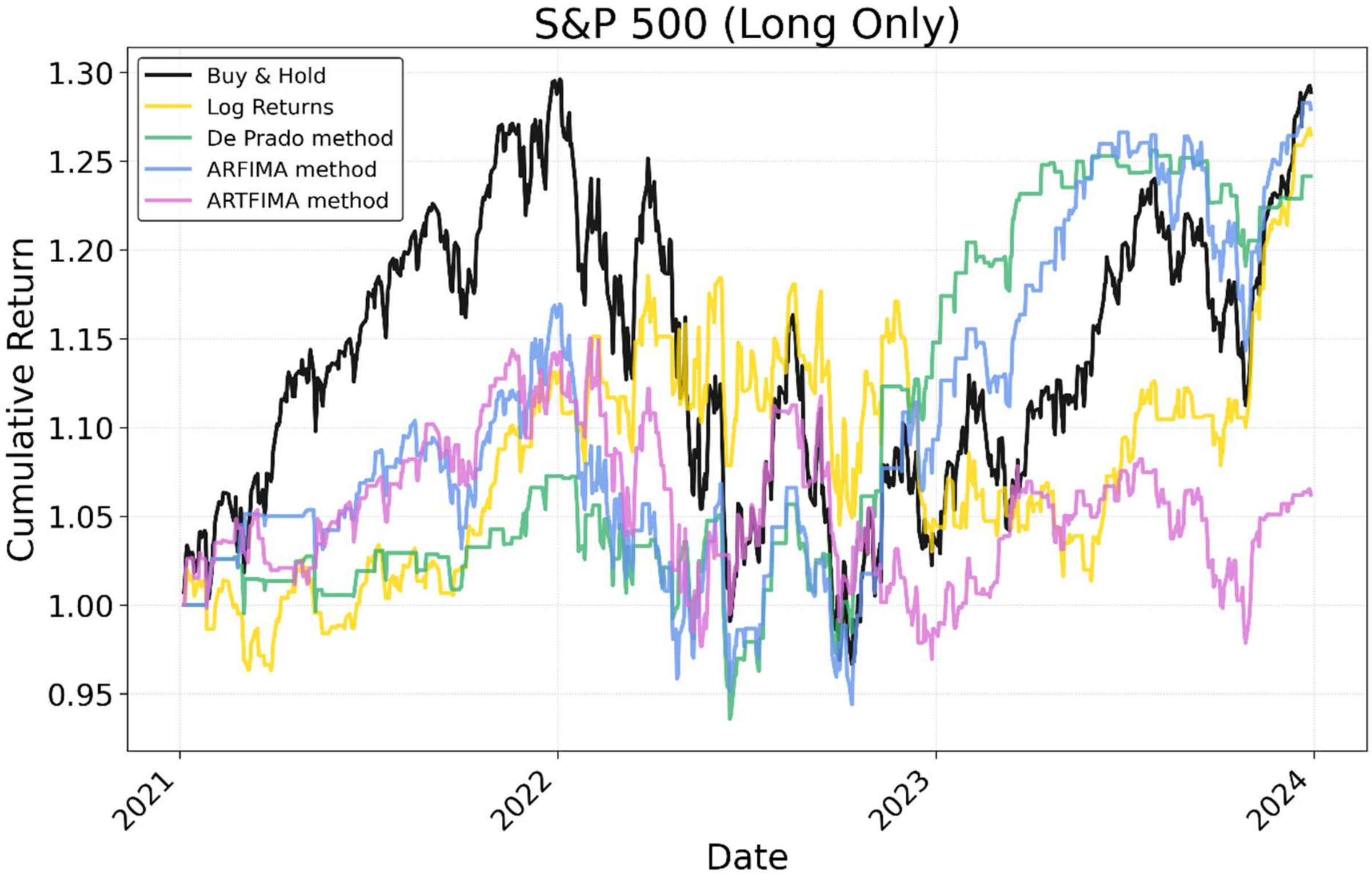

*Equity lines represent a trading performance for different techniques of data differentiation in period from 01.01.2021 to 31.12.2023. Transaction costs of 0.005% are included (Kryńska and Ślepaczuk, 2022).*

Source: Own elaborations

**Table 6.** Trading performance metrics for *Long Only* strategy, depending on different techniques of data differentiation for S&P 500

| Indicator | $\Delta X_t$ | $\Delta^{d_{De\ Prado}} X_t$ | $\Delta^{d_{ARFIMA}} X_t$ | $\Delta^{d_{ARTFIMA},\lambda_{ARTFIMA}} X_t$ | Buy & Hold |
|---|---|---|---|---|---|
| **ARC** | *8.14* | *7.48* | *8.56* | *2.03* | ***8.82*** |
| **ASD** | *14.01* | ***11.02*** | *13.72* | *13.88* | *17.59* |
| **MD** | *14.49* | ***12.75*** | *19.27* | *15.69* | *25.43* |
| **IR** | *0.58* | ***0.68*** | *0.62* | *0.15* | *0.50* |
| **SR** | *0.92* | ***1.10*** | *1.01* | *0.22* | *0.83* |

Source: Own elaborations

**Figure 14.** Equity lines for $Long - Short$ strategy, depending on different techniques of data differentiation for WIG20

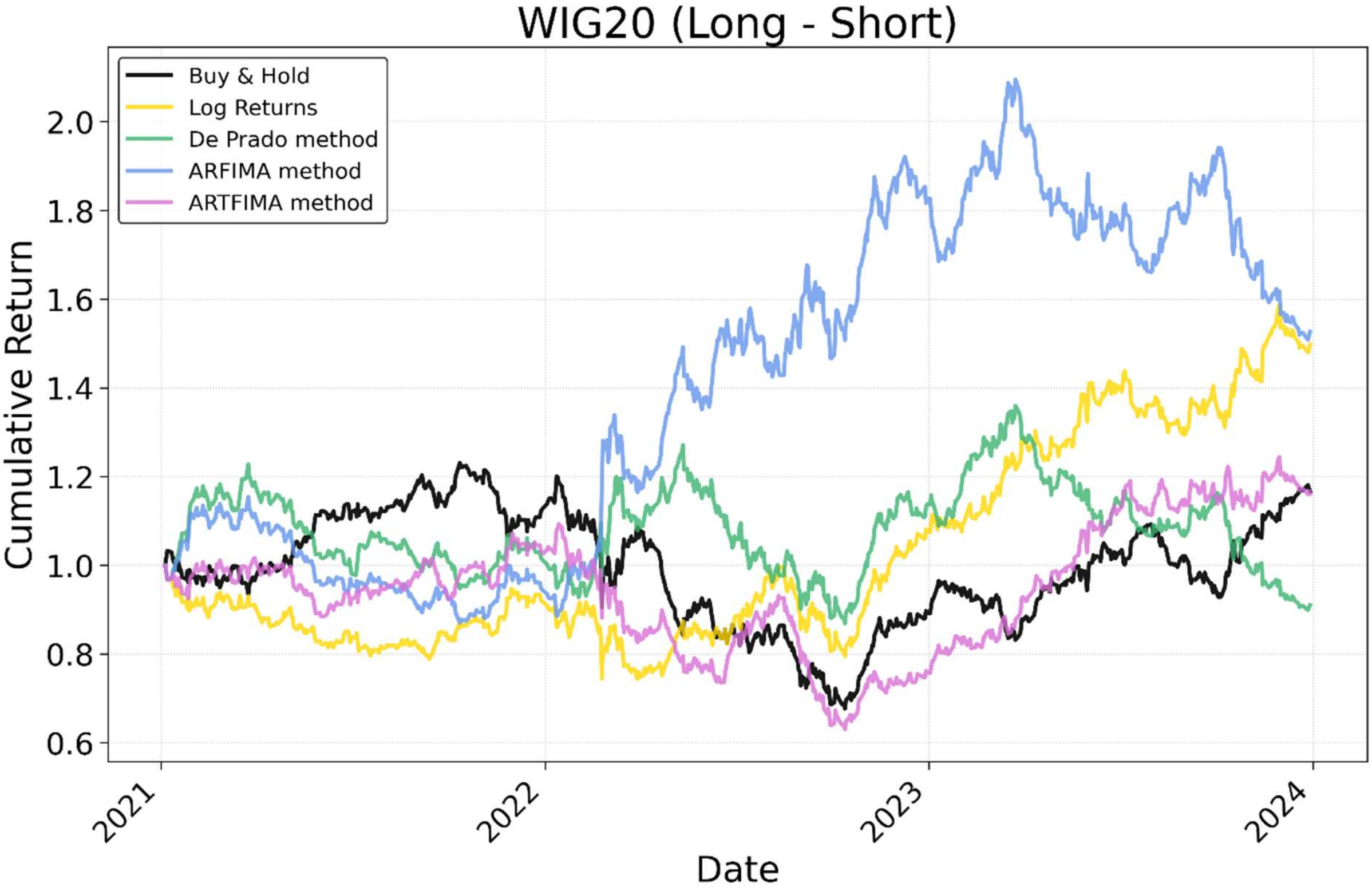

*Equity lines represent a trading performance for different techniques of data differentiation in period from 01.01.2021 to 31.12.2023. Transaction costs of 0.005% are included (Kryńska and Ślepaczuk, 2022).*

Source: Own elaborations

**Table 7.** Trading performance metrics for $Long - Short$ strategy, depending on different techniques of data differentiation for WIG20

| Indicator | $\Delta X_t$ | $\Delta^{d_{De\ Prado}} X_t$ | $\Delta^{d_{ARFIMA}} X_t$ | $\Delta^{d_{ARTFIMA}, \lambda_{ARTFIMA}} X_t$ | Buy & Hold |
|---|---|---|---|---|---|
| **ARC** | *14.40* | *-3.10* | ***15.14*** | *5.10* | *5.26* |
| **ASD** | ***23.75*** | *23.91* | *23.80* | *23.87* | *23.89* |
| **MD** | ***25.66*** | *33.87* | *28.01* | *42.34* | *45.05* |
| **IR** | *0.61* | *-0.13* | ***0.64*** | *0.21* | *0.22* |
| **SR** | *1.00* | *-0.21* | ***1.11*** | *0.34* | *0.37* |

Source: Own elaborations

**Figure 15.** Equity lines for $Long\ Only$ strategy, depending on different techniques of data differentiation for WIG20

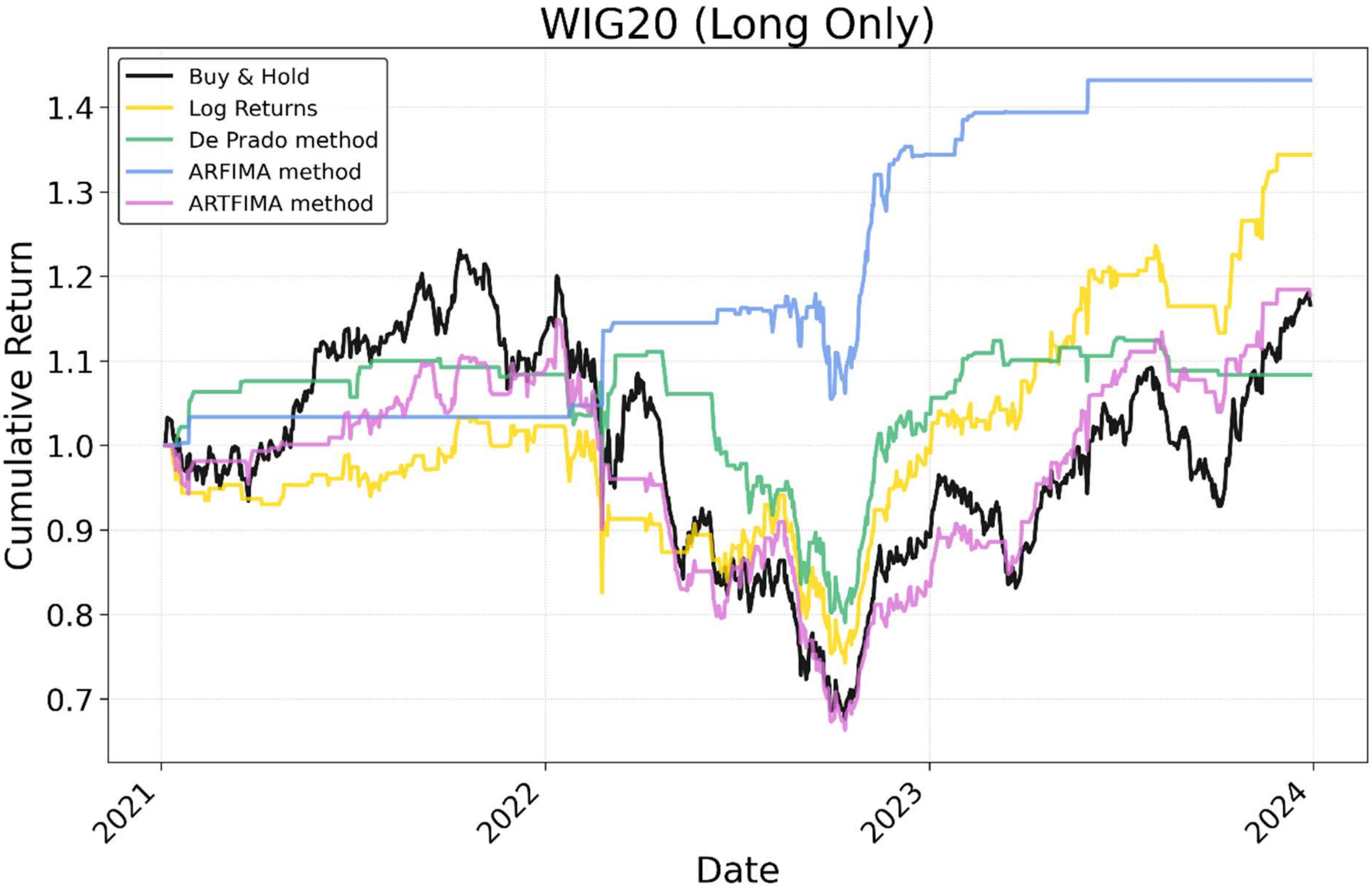

*Equity lines represent a trading performance for different techniques of data differentiation in period from 01.01.2021 to 31.12.2023. Transaction costs of 0.005% are included (Kryńska and Ślepaczuk, 2022).*

Source: Own elaborations

**Table 8.** Trading performance metrics for $Long\ Only$ strategy, depending on different techniques of data differentiation for WIG20

| Indicator | $\Delta X_t$ | $\Delta^{d_{De\ Prado}} X_t$ | $\Delta^{d_{ARFIMA}} X_t$ | $\Delta^{d_{ARTFIMA}, \lambda_{ARTFIMA}} X_t$ | Buy & Hold |
|---|---|---|---|---|---|
| **ARC** | *10.35* | *2.70* | ***12.71*** | *5.60* | *5.26* |
| **ASD** | *18.81* | *14.97* | ***9.89*** | *18.85* | *23.89* |
| **MD** | *28.04* | *28.85* | ***10.56*** | *42.24* | *45.05* |
| **IR** | *0.55* | *0.18* | ***1.28*** | *0.30* | *0.22* |
| **SR** | *0.85* | *0.26* | ***2.47*** | *0.45* | *0.37* |

Source: Own elaborations

**Figure 16.** Equity lines for $Long - Short$ strategy, depending on different techniques of data differentiation for DAX

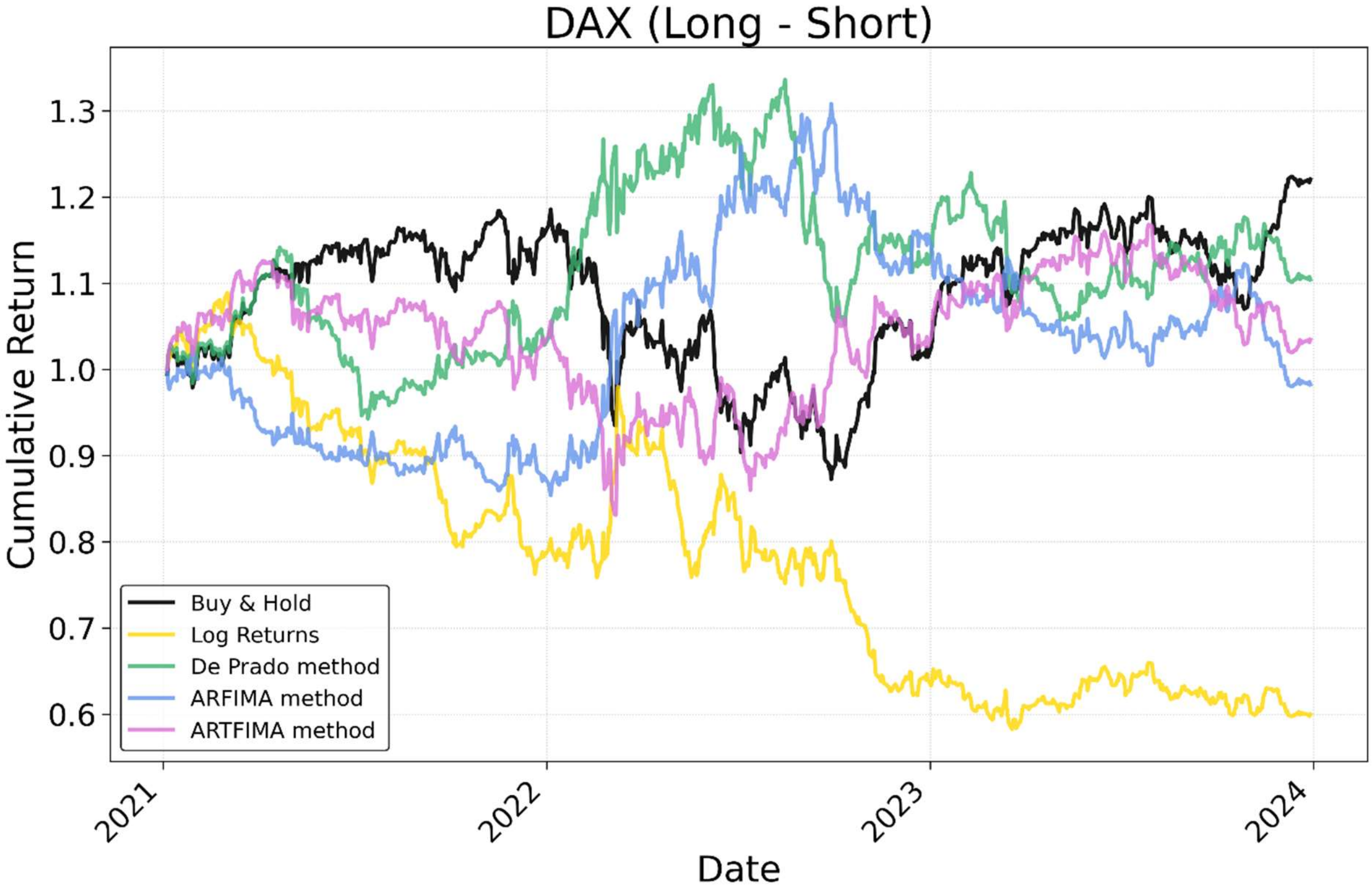

*Equity lines represent a trading performance for different techniques of data differentiation in period from 01.01.2021 to 31.12.2023. Transaction costs of 0.005% are included (Kryńska and Ślepaczuk, 2022).*

Source: Own elaborations

**Table 9.** Trading performance metrics for $Long - Short$ strategy, depending on different techniques of data differentiation for DAX

| Indicator | $\Delta X_t$ | $\Delta^{d_{De\,Prado}} X_t$ | $\Delta^{d_{ARFIMA}} X_t$ | $\Delta^{d_{ARTFIMA}, \lambda_{ARTFIMA}} X_t$ | Buy & Hold |
|---|---|---|---|---|---|
| **ARC** | *-15.68* | *3.35* | *-0.59* | *1.13* | ***6.86*** |
| **ASD** | *17.48* | ***17.45*** | *17.48* | *17.47* | *17.51* |
| **MD** | *46.48* | ***21.28*** | *25.07* | *26.19* | *26.40* |
| **IR** | *-0.90* | *0.19* | *-0.03* | *0.06* | ***0.39*** |
| **SR** | *-1.49* | *0.31* | *-0.06* | *0.11* | ***0.64*** |

Source: Own elaborations

**Figure 17.** Equity lines for *Long Only* strategy, depending on different techniques of data differentiation for DAX

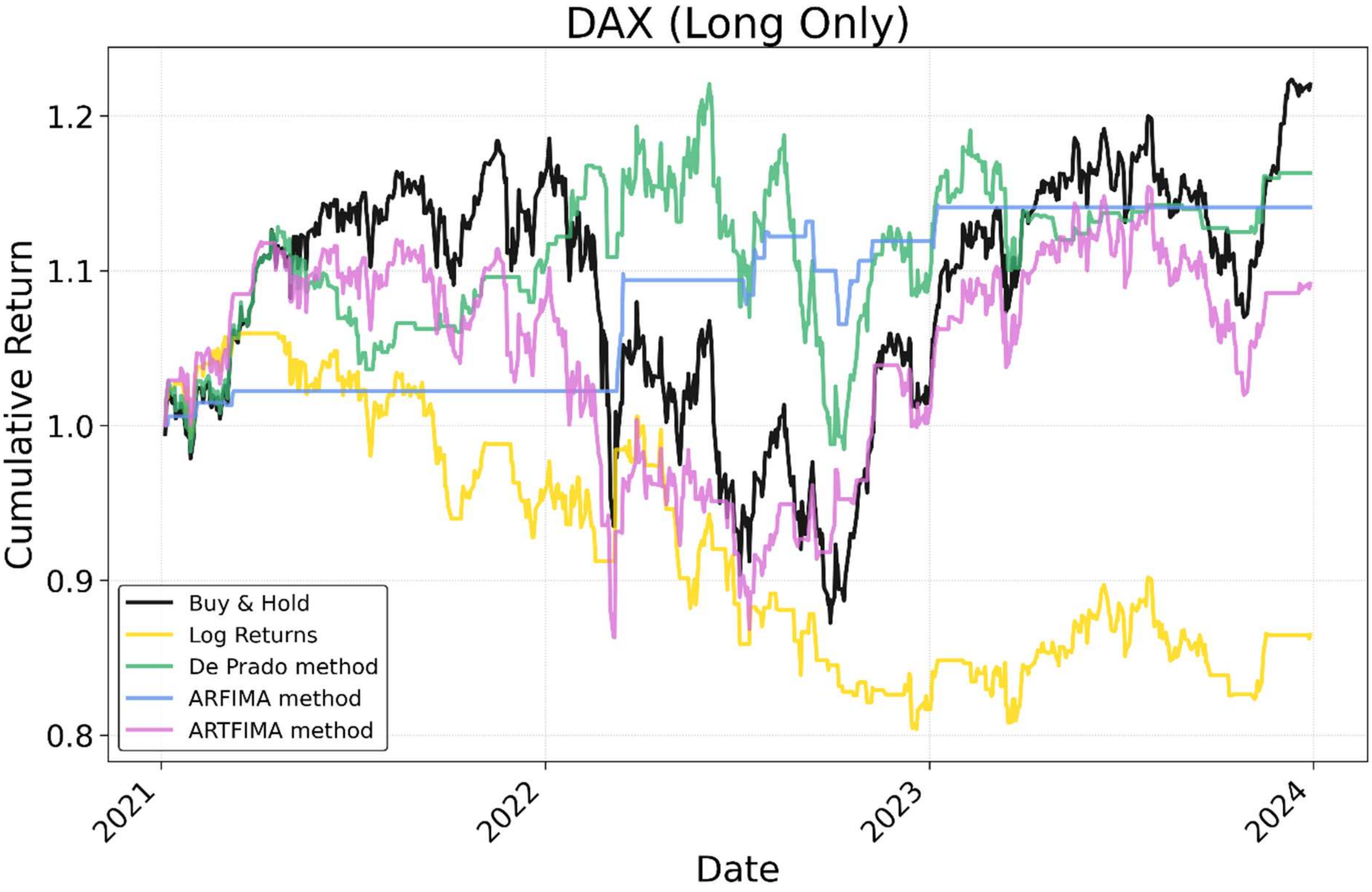

*Equity lines represent a trading performance for different techniques of data differentiation in period from 01.01.2021 to 31.12.2023. Transaction costs of 0.005% are included (Kryńska and Ślepaczuk, 2022).*

Source: Own elaborations

**Table 10.** Trading performance metrics for *Long Only* strategy, depending on different techniques of data differentiation for DAX

| Indicator | $\Delta X_t$ | $\Delta^{d_{De\ Prado}} X_t$ | $\Delta^{d_{ARFIMA}} X_t$ | $\Delta^{d_{ARTFIMA}, \lambda_{ARTFIMA}} X_t$ | Buy & Hold |
|---|---|---|---|---|---|
| **ARC** | *-4.73* | *5.17* | *4.49* | *2.97* | ***6.86*** |
| **ASD** | *12.52* | *12.54* | ***4.17*** | *15.47* | *17.51* |
| **MD** | *24.12* | *19.33* | ***5.87*** | *22.94* | *26.40* |
| **IR** | *-0.38* | *0.41* | ***1.08*** | *0.19* | *0.39* |
| **SR** | *-0.58* | *0.63* | ***2.40*** | *0.31* | *0.64* |

Source: Own elaborations

**Figure 18.** Equity lines for $Long - Short$ strategy, depending on different techniques of data differentiation for Nikkei 225

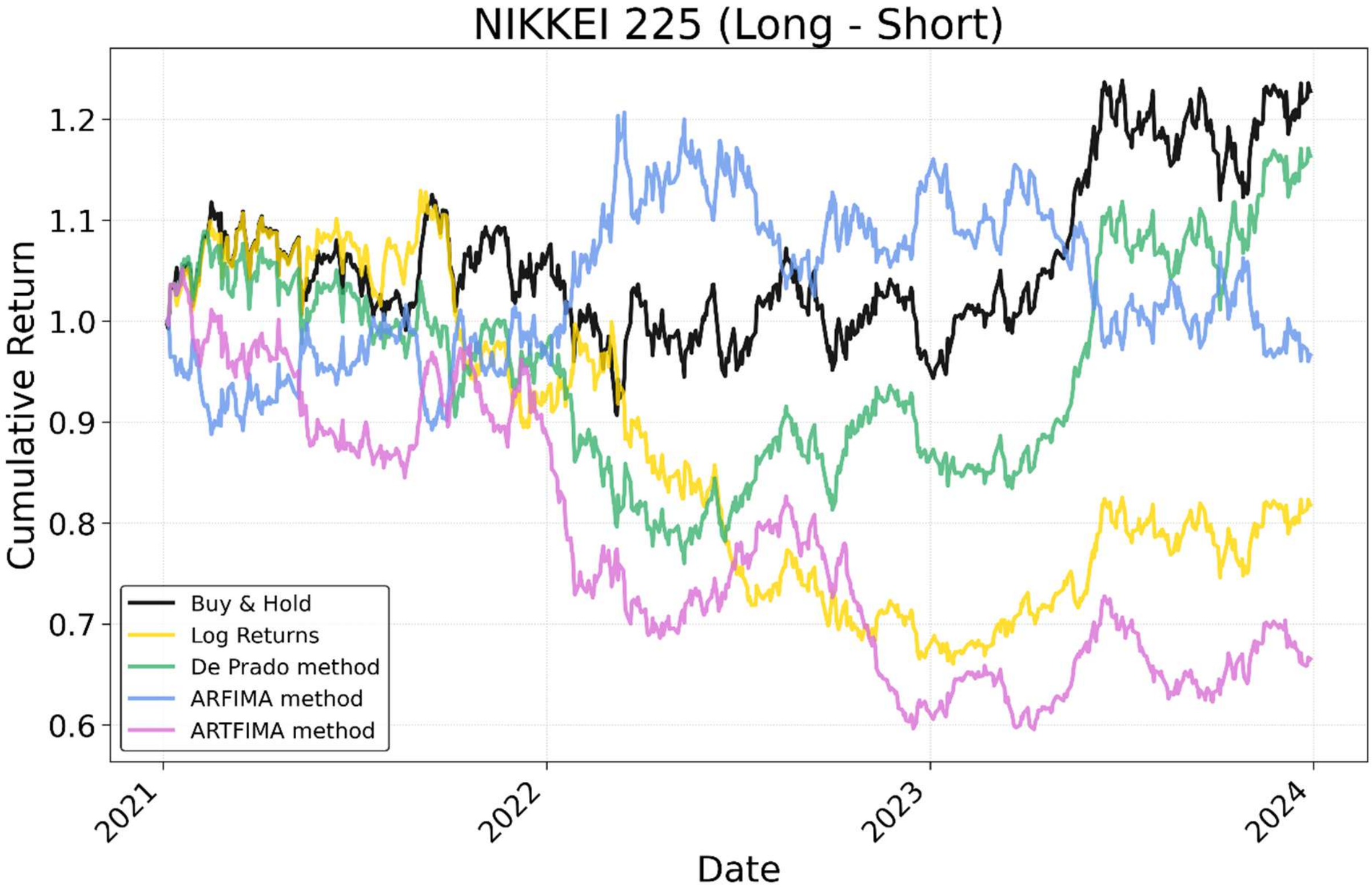

*Equity lines represent a trading performance for different techniques of data differentiation in period from 01.01.2021 to 31.12.2023. Transaction costs of 0.005% are included (Kryńska and Ślepaczuk, 2022).*

Source: Own elaborations

**Table 11.** Trading performance metrics for $Long - Short$ strategy, depending on different techniques of data differentiation for Nikkei 225

| Indicator | $\Delta X_t$ | $\Delta^{d_{De\ Prado}} X_t$ | $\Delta^{d_{ARFIMA}} X_t$ | $\Delta^{d_{ARTFIMA}, \lambda_{ARTFIMA}} X_t$ | Buy & Hold |
|---|---|---|---|---|---|
| **ARC** | *-6.48* | *5.18* | *-1.13* | *-12.70* | ***7.08*** |
| **ASD** | ***18.35*** | *18.45* | *18.38* | *18.43* | *18.45* |
| **MD** | *41.51* | *30.20* | *20.43* | *43.39* | ***19.41*** |
| **IR** | *-0.35* | *0.28* | *-0.06* | *-0.69* | ***0.38*** |
| **SR** | *-0.58* | *0.47* | *-0.11* | *-1.12* | ***0.65*** |

Source: Own elaborations

**Figure 19.** Equity lines for *Long Only* strategy, depending on different techniques of data differentiation for Nikkei 225

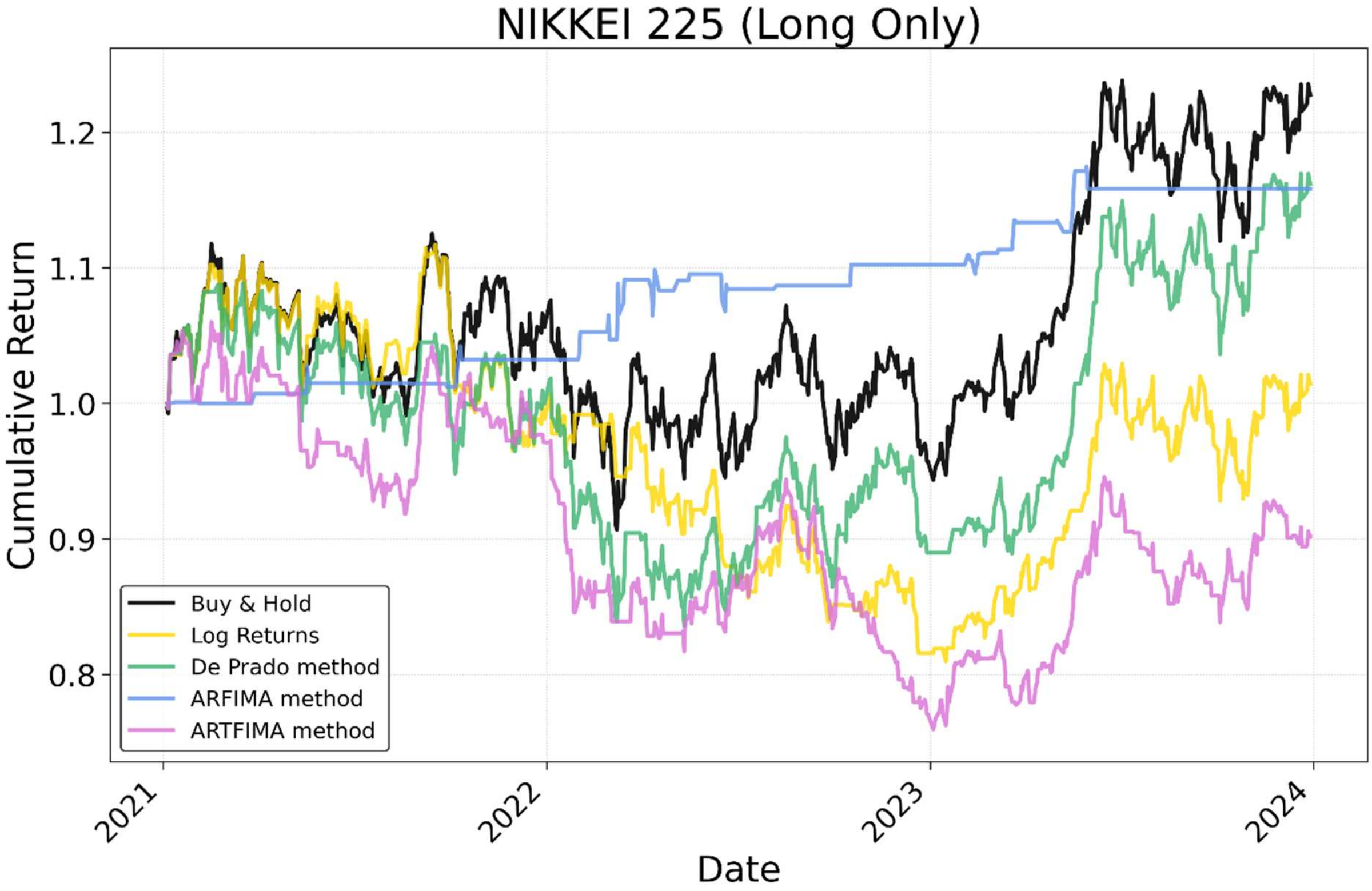

*Equity lines represent a trading performance for different techniques of data differentiation in period from 01.01.2021 to 31.12.2023. Transaction costs of 0.005% are included (Kryńska and Ślepaczuk, 2022).*

Source: Own elaborations

**Table 12.** Trading performance metrics for *Long Only* strategy, depending on different techniques of data differentiation for Nikkei 225

| Indicator | $\Delta X_t$ | $\Delta^{d_{De\ Prado}} X_t$ | $\Delta^{d_{ARFIMA}} X_t$ | $\Delta^{d_{ARTFIMA}, \lambda_{ARTFIMA}} X_t$ | Buy & Hold |
|---|---|---|---|---|---|
| **ARC** | *0.48* | *5.13* | *5.02* | *-3.38* | ***7.08*** |
| **ASD** | *15.99* | *17.57* | ***5.13*** | *14.53* | *18.45* |
| **MD** | *27.52* | *23.21* | ***2.79*** | *28.35* | *19.41* |
| **IR** | *0.03* | *0.29* | ***0.98*** | *-0.23* | *0.38* |
| **SR** | *0.05* | *0.48* | ***1.80*** | *-0.36* | *0.65* |

Source: Own elaborations